# AN ASSESSMENT OF SUNSPOT NUMBER DATA COMPOSITES OVER 1845−2014

M. LOCKWOOD [1], M.J. OWENS [1], L. BARNARD [1], AND I.G. USOSKIN [2,3]

[1] Department of Meteorology, University of Reading, Earley Gate, Reading, RG6 6BB, UK
m.lockwood@reading.ac.uk
[2] ReSoLVE Centre of Excellence, P.O.Box 3000, FIN-90014 University of Oulu, Finland
[3] also at Sodankyla Geophysical Observatory, Oulu, Finland
*Received x, accepted y, published z*

ABSTRACT

New sunspot data composites, some of which are radically different in the character of their long-term variation, are evaluated over the interval 1845−2014. The method commonly used to calibrate historic sunspot data, relative to modern-day data, is "daisy-chaining", whereby calibration is passed from one data subset to the neighbouring one, usually using regressions of the data subsets for the intervals of their overlap. Recent studies have illustrated serious pitfalls in these regressions and the resulting errors can be compounded by their repeated use as the data sequence is extended back in time. Hence the recent composite data series by Usoskin et al. (2016), $R_{UEA}$, is a very important advance because it avoids regressions, daisy-chaining and other common, but invalid, assumptions: this is achieved by comparing the statistics of "active day" fractions to those for a single reference dataset. We study six sunspot data series including $R_{UEA}$ and the new "backbone" data series ($R_{BB}$, recently generated by Svalgaard and Schatten, (2016) by employing both regression and daisy-chaining). We show that all six can be used with a continuity model to reproduce the main features of the open solar flux variation for 1845−2014, as reconstructed from geomagnetic activity data. However, some differences can be identified that are consistent with tests using a basket of other proxies for solar magnetic fields. Using data from a variety of sunspot observers, we illustrate problems with the method employed in generating $R_{BB}$ which cause it to increasingly overestimate sunspot numbers going back in time and we recommend using $R_{UEA}$ because it employs more robust procedures that avoid such problems.

1. INTRODUCTION

Sunspot number is a primary index of long-term solar activity (Usoskin, 2013; Hathaway, 2015) and its reliable definition is of importance for studies of the solar dynamo, solar irradiance, coronal physics, space weather, space climate, and solar-terrestrial relations. The sunspot number is defined daily by the formula introduced by Wolf (1861):

$$R = k \times (10 N_G + N_S) \qquad (1)$$

where $N_G$ is the number of sunspot groups, $N_S$ is the number of individual sunspots, and $k$ is a calibration factor that varies with location, instrumentation and observer procedures. Before 1982, compilation of $R$ used a single primary observer for most days (on some days after 1877 when no primary observer could make observations, an average from secondary observers was used); after 1982 multiple observers on each day were used. The $k$ factors for different observers can differ by a factor as large as three (Clette *et al.*, 2015) and so are critical to the accurate quantification of $R$. To extend sunspot data to times before when both $N_G$ and $N_S$ were recorded systematically, Hoyt et al. (1994) and Hoyt and Schatten (1998) defined the group sunspot number $R_G$ to be

$$R_G = 12.08 < k' \times N_G >_n \qquad (2)$$

where $k'$ is the site/observer calibration factor for sunspot groups only and the averaging is carried out over the $n$ observers who are available for the day in question. The factor of 12.08 makes the means of $R_G$ and $R$ (specifically, version 1 of the international sunspot number, $R_{ISNv1}$, see below) the same over 1875−1976. Note that assuming that the $k$ or $k'$ factors in equations (1) and (2) are constants assumes that the counts from different observers are proportional to each other, such that application of the appropriate constant multiplicative factor renders them the same. Initially, Wolf considered that the $k$ factors were constant for each observer (Wolf, 1861) but he later realised that this was not, in general, valid and that observer's $k$ and $k'$ factors depend on the level of solar activity (Wolf 1873) and so they were calculated either quarterly or annually (using daily data) at the Zürich observatory (see Friedli, 2016). It is well known that estimates of $R$ and $R_G$ diverge as one goes back in time. This could be due to real long-term changes in the ratio $N_S/N_G$, but otherwise it would reflect erroneous long-term drifts in the calibration factors for either $R$ or $R_G$ (i.e., $k$ and $k'$, respectively) or both. Recently, Friedli (2016) has shown that the ratio $N_S/N_G$ has a regular solar cycle variation but no long term change and so can be used as a way of calibrating different observers. A series of workshops were held in recent years to try to investigate the differences between $R$ and $R_G$ (Clette et al., 2015). This has stimulated the generation of a number of new sunspot number and sunspot group number composites. These vary in a surprisingly radical way with considerable implications for our understanding of the solar dynamo and its variability. The methods used to make these sunspot number data composites, and the centennial-scale variations in the derived data series, are reviewed and assessed in this paper.

Both sunspot numbers and sunspot group numbers are synthetic indices and somewhat limited indicators of solar magnetic activity. They give information on the larger magnetic features in the photosphere only and they do not vary linearly with many of the key parameters of solar and heliospheric activity and structure. Moreover, there is a threshold effect whereby a lack of sunspots does not necessarily imply the absence of the cyclic solar activity. For recent solar cycles we have other metrics that are more directly relevant and measured with less subjectivity: as a result, sunspot numbers are of importance mainly because of the longevity of the data sequence. Hence if sunspots numbers are to be useful, it is vital check that their long-term variation is as accurately reproduced as it can be. That is the aim of the present paper.

The key problem in generating homogeneous composites of $R$ and $R_G$ is the estimation of the $k$ and $k'$ factors for the historic observers. Until recently, all composites used "daisy-chaining" whereby the calibration is passed from the data from one observer to that from the previous or next observer (depending on whether the compiler is working, respectively, backwards or forwards in time) by comparison of data during an overlap period when both made observations. Hence, for example, if proportionality is assumed and intercalibration of observer numbers $i$ and $(i+1)$ in the data composite yields $k_i/k_{i+1} = f_i^{(i+1)}$ then daisy chaining means that the first ($i = 1$) and last ($i = n$) observer's $k$ factors are related by $k_1 = k_n \prod_{i=1}^{n}(f_i^{(i+1)})$. A similar product applies for the $k'$ factors for group sunspot numbers. Hence daisy chaining means that all sunspot and sunspot group numbers, relative to modern values, are influenced by all of the intercalibrations between data subsets at subsequent times.

Because meteorological conditions vary with location and from day-to-day, and some sunspot groups last for only one day (Willis et al., 2016), it is important to compare observers only on a daily basis and only on days when both were able to make observations. Otherwise, significant errors are caused by days when observations were not possible if annual or monthly means are compared. Often comparisons have been made using linear, ordinary least squares regression. Errors caused by inadequate and/or inappropriate regression techniques were discussed by Lockwood et al. (2006) in relation to differences between reconstructions of the magnetic field in near-Earth interplanetary space from geomagnetic activity data. The seriousness of potential problems has been expressed succinctly by Nau (2016): "If any of the assumptions is violated (i.e., if there are nonlinear relationships between dependent and independent variables or the errors exhibit correlation, heteroscedasticity, or non-normality), then the forecasts, confidence intervals, and scientific insights yielded by a regression model may be (at best) inefficient or (at worst) seriously biased or misleading." Lockwood et al. (2016c) have studied these pitfalls in the context of sunspot group numbers, using annual means of observations from the Royal Observatory, Greenwich / Royal Greenwich Observatory (hereafter "RGO") for after 1920, when there are no concerns about their calibration. They compared the RGO sunspot group numbers with data synthesised to simulate what a lower-acuity observer (i.e., one who has a higher $k'$) would have seen. This was done by assuming the lower acuity observer would only detect groups above a threshold of total spot area in the group (uncorrected for foreshortening near the limb, i.e. as detected by the observer) and studying the effect of this threshold. It was shown that there is no single regression procedure that always retrieves the original RGO data and tests must be applied to check that the assumptions inherent in the procedure applied are not violated. Specifically, it was shown that errors of up to 30% could arise in one regression of annual mean data even for two data series with a correlation exceeding 0.98 over two full solar cycles. The biggest problems are associated with non-linearity and non-normal distributions of data errors which violate the assumptions made by most regression techniques: such errors should always be tested for (for example using a quantile-quantile ("Q-Q") plot comparison against a normal distribution) before a correlation is used for any scientific inference or prediction (Lockwood et al., 2006, 2016c).

Lockwood et al. (2016c) confirmed that significant errors were introduced by assuming proportionality between the results of two observers and that this is, in both principle and practice, incorrect and leads to non-normal error distributions and hence errors in regressions. In fact, sets of sunspot data often do not have a linear relationship. Using the ratio of sunspot numbers (or sunspot group numbers) from two different observers also implicitly assumes proportionality and generates asymmetric errors that vary hyperbolically with the denominator, such that both the ratio and its uncertainty tend to infinity as the denominator tends to zero. This has been dealt with in two ways in the past: (1) neglecting values where the denominator falls below an arbitrarily-chosen threshold; and (2) taking averages over an extended period (greater than a solar cycle) so the denominator does not become small. Neither of these is satisfactory: on top of generating asymmetric error distributions, method (1) preferentially removes solar minimum values and method (2) matches the mean values but loses information about the solar cycle amplitudes because sunspot numbers and sunspot group numbers do not fall to zero in all minima. It is not necessary to assume proportionality (or even linearity), nor to make use of ratios, nor to ignore the effect of missing observation days. Hence adherence to good practice can avoid all of the associated pitfalls. Unfortunately, some reconstructions make use of one of more of these unreliable practices and it is easy, but not satisfactory, to dismiss without proof the effects of this as being small.

These issues are particularly important in daisy-chaining of calibrations to generate a long-interval data composite because errors compound with successive regressions (Lockwood et al., 2016b, c). For these reasons, the recent group sunspot number reconstruction by Usoskin et al. (2016) is a very important development because it avoids using either regression or daisy-chaining and does not even need to assume that the $k'$ factors (for a given level of solar activity) remained constant for any one observer (although, for simplicity, this assumption was made in the initial paper). In addition, the method assumes neither proportionality nor linearity between the results of different observers and evaluates each observer on a daily basis and not using monthly or annual means. This rigour was achieved by comparing all data to a standard dataset covering a reference period (the RGO data between 1900 and 1976 were used, and this standard is evaluated in section 2.3). This means that, for example, isolated fragments of data, disconnected from the data sequence by a data gap, can be employed without having to use questionable data, or an assumption, to bridge that gap – something that cannot be done for any form of daisy-chaining. Furthermore, should any segment of data be incorrect or badly calibrated, the error does not corrupt any other data segments, whereas for daisy-chaining the error propagates from that segment to all others calibrated from it: thus every error infects all prior data (if the calibrations are passed back in time, starting from modern data) and if they arise from the systematic application of unreliable procedures, these errors will compound.

In the Usoskin et al. (2016) procedure, the comparisons with the reference dataset are made by, effectively, considering the relationship between an observer's sunspot group count and the statistics of the fraction of all observation days that were "active" (i.e., on which sunspots were observed). Hence the only requirement is that the observer had distinguished between days on which he/she could see the solar disk but detected no sunspots (i.e. non-active days) from days on which the solar disk could not be observed (i.e., missing data, for example due to cloud cover). The method uses the probability distribution functions (pdfs) of different group numbers and makes no assumptions of proportionality or linearity of the relationship between the data from different observers.

Another example of the use of a non-parametric, daisy-chain-free calibration of observers is the recent work by Friedli (2016) who re-calibrated observers using the statistics of $N_S$, $N_G$ and the ratio $N_S/N_G$. At the time of writing, this work has yet to be published so we do not include it here as one of the data composites tested: however, the data sequence derived by Friedli (2016) is similar to $R_{UEA}$ which is tested.

## 2. SUNSPOT DATA COMPOSITES

We here study six different sunspot number and sunspot group number data composites, introduced in the following six sub-sections. These are plotted in the six panels of figure 1 and, to enable comparisons, each is compared to the same black line which is the median $R_{med}$ of all available sequences for each year (which number three in 1650, rising to six by the present day). To compute $R_{med}$, all group numbers have been multiplied by the 12.08 normalisation factor adopted by Hoyt and Schatten (1998) for $R_G$ (see equation 2).

### 2.1 The International Sunspot Number Version 1, $R_{ISNv1}$

This is a composite of sunspot numbers, as defined by equation (1), initially generated by Wolf and continued at the Zürich observatory until 1980 and subsequently compiled by SIDC (the World Data Center for the production, preservation and dissemination of the international sunspot number and the Solar Physics Research department of the Royal Observatory of Belgium) until July 2015 when it was replaced by version 2 (see section 2.2). Like all the series, except that by Usoskin et al. (2016) (see section 2.4), the calibration is by daisy-chaining. The annual means are shown by the brown line in figure 1(f-i), while figure 1(f-ii) shows the difference between $R_{ISNv1}$ and $R_{med}$. $R_{ISNv1}$ covers the interval 1818–2014. The primary station, assumed to have $k = 1$, was Zürich until 1980 after which the Specola Solare Observatory in Locarno was used as the standard.

**2.2 The International Sunspot Number Version 2, $R_{ISNv2}$**

In July 2015, SIDC changed its primary data product to $R_{ISNv2}$, in which many data were re-calibrated to make a number of corrections to $R_{ISNv1}$ (Clette et al., 2015). It should be noted that this series must be scaled down by a factor 0.6 to be compared to $R_{ISNv1}$ because it was decided to dispense with a factor that had been applied in the generation of $R_{ISNv1}$ for historical reasons. The most recent correction is to allow for a drift in the Locarno reference station $k$ value. This drift was found by research aimed at explaining why the relationship between the F10.7 solar radio flux and $R_{ISNv1}$ broke down dramatically just after the long and low activity minimum between solar cycles 23 and 24 (Johnson, 2011). The Locarno $k$-values were re-assessed using the average of sixteen other stations (out of a total of about eighty) that provided near-continuous data over the 32-year interval studied. The results showed that the Locarno $k$-factors had varied between 1.15 in 1987 and 0.85 in 2009 (i.e. by ±15%). The best evidence for this correction is the large number of sunspot observers that vary in the same way with respect to the Locarno data, but we also note that it is also supported by tests against ionospheric data (Lockwood et al., 2016a).

A second major correction is for what has become termed the "Waldmeier discontinuity". (Svalgaard, 2011; Aparicio *et al.*, 2012, Cliver *et al.*, 2013). This is thought to have been caused by the introduction of a weighting scheme for sunspot counts according to their size and a change in the procedure used to define a group (including the so-called "evolutionary" classification that considers how groups evolve from one day to the next); both changes that may have been introduced by the then director of the Zürich observatory, Max Waldmeier, (Hockey, 2014) after he took over responsibility for the production of the Wolf sunspot number in 1945. Note that these changes affect the sunspot numbers and the sunspot group numbers used to derive them in Zürich, but not necessarily by the same amount. Note also that this discontinuity affects only Zürich data (and datasets calibrated to it) but is not relevant to independent data such the data generated at RGO. However, as discussed by Friedli (2016), some of these changes might have been made gradually since the group number weighting was partly used by other observers (e.g., Wolfer, Brunner) before Waldmeier took charge of the Zürich observatory in 1945.

The changes made by Waldmeier improved the sunspot number as a metric of solar magnetic activity and gave an algorithm that was improved, fixed and better defined. However, Waldmeier would have been unable to apply his new algorithm to much of the prior data retrospectively and so it was inevitable that his improvements led to a discontinuity, of some magnitude, in the composite series. Note that the only options open to Waldmeier were either to improve the metric or to knowingly continue to use less-than-optimal existing procedures to remain fully compatible with prior data. From a modern perspective, it is easy to think that Waldmeier made the wrong choice as we now have other, more specific and less subjective solar metrics and observations and we use sunspots mainly to understand long-term variations. However, in 1945 priorities were different because relationships between sunspots and factors such as ionospheric plasma concentrations were being discovered and explored and hence the requirement was to make sunspot numbers as accurate and representative of solar activity as they could be. Hence Waldmeier made a decision that was appropriate to the science of his day.

By comparison with other long time-series of solar and solar-terrestrial indices, Svalgaard (2011) makes a compelling case that this discontinuity is indeed present in the Zürich data series at about 1945. However, there is debate as to how large the correction should be, debate that is discussed in section 3 of the present paper. There is also debate as to whether or not the correction is a simple multiplicative factor (i.e. the corrected data should be proportional to the uncorrected data and the discontinuity is just a sensitivity change, making the corrected sunspot number $R' = f_R R$) or if there is also effectively a zero level offset ($R' = f_R R + \delta$) or indeed is it non-linear, such that the effect at high and low solar activity is different ($R' = f_R R^n + \delta$). The $R_{ISNv2}$ series assumes proportionality and employs a multiplicative factor of $f_R = 1.18$, i.e. values before the discontinuity need increasing by 18% to become consistent with modern values (Clette and Lefèvre, 2016).

There are other calibration debates inherent in $R_{ISNv1}$. For example, Leussu (2013) studied the difference between the data of Schwabe and of Wolf and concluded that $R_{ISNv1}$ should be reduced by 20% before 1848. This conclusion is contested by Clette et al (2015). As this only influences the first 3 years of the interval studied here, this issue is not considered further in the current paper. Another debated inter-calibration is between the data generated by Schwabe (which ends in 1867) and by Wolfer (which commences a whole solar cycle later in 1878). This is addressed in section 5 of the present paper.

The variation of $R_{ISNv2}$ is shown in mauve in Figure 1(e-i) and its deviation from the median $R_{med}$ in figure 1(e-ii).

**2.3 The Group Sunspot Number of Hoyt and Schatten, $R_G$**

The group sunspot number, as defined by equation (2), was introduced by Hoyt et al. (1994) and Hoyt and Schatten (1998) who generated an intercalibrated series that begins in 1610 and has been much used. For 1875–1976, $R_G$ uses the RGO photo-heliographic sunspot group data (Willis et al., 2013a, 2013b). This has been updated to the present day using the group sunspot data generated by the SOON network as the RGO observations ceased in 1976. The version shown in green in Figure 1(d-i) uses the calibration of RGO and SOON data, derived by two different statistical techniques by Lockwood et al., (2014a). It also employs some corrections to the 17$^{th}$ century data by Vaquero et al. (2011): its deviation from the median $R_{med}$ is shown in figure 1(d-ii). With the SOON data added, $R_G$ extends from 1610 to the present day.

The RGO data, and hence $R_G$, are fully independent of $R_{ISNv1}$ (using different observations, scaling practices and personnel) and are not influenced in any way by the Waldmeier discontinuity. Indeed, for 1918-1976 $R_G$ provides a valuable standard for comparisons because, uniquely, it can be reproduced because the original RGO photographic plates have survived. These raw data can be re-analysed to check the stability of the $k'$ factors in the work of the RGO observers who made the sunspot group counts. The plates have been digitized by the Mullard Space Science Laboratory in the UK and analysed with an automated scaling algorithm which can derive sunspot group areas and numbers (Çakmak, 2014). This automated scaling of the RGO images reproduces the manually scaled daily sunspot-group numbers well, with a correlation of monthly values of over 0.93; however, there are differences, as discussed below and demonstrated by the annual means shown in figure 2 (from Tlatov and Ershov, 2014).

Lockwood et al. (2016c) compared RGO data with deliberately-degraded RGO data to demonstrate that the relationship between observers of different visual acuities is, in general, non-linear. Figure 2 demonstrates the good agreement between the RGO dataset and other data, once this non-linearity is accounted for. Parts (a) and (b) of this figure compare annual-mean group number data from the standard RGO dataset ($[N_G]_{RGO}$, in black) with that from Mount Wilson Observatory ($[N_G]_{MWO}$, in blue), from the Solar Observatory of the National Astronomical Observatory of Japan ($[N_G]_{NAOJ}$, in green), and the auto-scaled data from the RGO photoheliographic plates ($[N_G]_{RGO2}$, in red). The MWO data are often given as the number of independent groups in 10 month intervals and have been re-calculated here to be annual means of daily $N_G$, as for the other data. In figure 2(a) the data have been scaled linearly over the interval 1920-1945. It can be seen that agreement over this interval is very good but that this linear scaling leads to a peak of $[N_G]_{MWO}$ in solar cycle 19 (around 1958) that is larger than the peak in $[N_G]_{RGO}$ and much larger in the auto-scaled RGO data, $[N_G]_{RGO2}$. This non-linearity is investigated in parts (c), (d) and (e) of figure 2. Figure 2(c) is a scatter plot $[N_G]_{NAOJ}$ as a function of $[N_G]_{RGO}$. A linear fit of the RGO and NAOJ data over the full period of their overlap gives an overall $k'$ value of 1.050, if the RGO data are taken to define $k' = 1$. For these data, the plot is close to linear and the best fit regression line shown passes through the origin. Hence in this case, the RGO and NAOJ data are similar enough that proportionality does apply. For the MWO data, the corresponding $k'$ value is 0.916, and the relationship has become slightly nonlinear. The line is the best-fit $2^{nd}$-order polynomial. Note that the regression no longer passes through the origin but MWO is detecting spots at some of the times when RGO is not; i.e., $[N_G]_{MWO} > 0$ when $[N_G]_{RGO} = 0$, consistent with MWO being a higher-acuity observer than RGO (see Lockwood et al., 2016c). This is even more apparent for the rescaled RGO data which finds more groups than the original scaling of the RGO data ($k' = 0.882$) because it uses a less conservative definition of what constitutes a sunspot group. Both the non-linearity and the non-zero intercept are even more pronounced in this case. Taking the $2^{nd}$-order polynomial scaling gives the variations shown in figure 2(b). It can be seen that allowing for the non-linearity makes the variations of all these datasets very similar to the original RGO data. This highlights the importance of allowing for the non-linearity of the relationship of data from different acuity observers. For weaker solar cycles, linearity is a good approximation, but figure 2 shows that, for example, the peak of cycle 19 is, relatively, much greater for high-acuity observers than for lower-acuity ones because of the non-linear effect.

However, it has been suggested that the RGO data suffer from a data-quality problem before 1885 (Clette et al., 2014; Cliver and Ling, 2016): this cannot be verified or disproved in the same manner because the photographic RGO plates before 1918 have been lost (thought to have been destroyed during World War I). Because calibrations were daisy-chained by Hoyt and Schatten (1998), such an error would influence all earlier values of $R_G$.

### 2.4 The Group Sunspot Number of Usoskin et al., $R_{UEA}$

As discussed above, this reconstruction is the only one to avoid using both daisy-chaining and regressions. Because the standard used to calibrate all data is the RGO data for 1900–1976, and because the SOON data are added to the RGO data using the intercalibration of Lockwood et al. (2014), $R_{UEA}$ is the same as $R_G$ after 1900. Note that $R_{UEA}$, like $R_G$, has no correction for the Waldmeier discontinuity, nor should it as it is not influenced by any of the factors that gave rise to that putative discontinuity. The variation of $R_{UEA}$ is shown in orange in Figure 1(c-i) and its deviation from the median $R_{med}$ in figure 1(c-ii).

### 2.5 The "Backbone" Group Sunspot Number, $R_{BB}$

Another new group number reconstruction has recently published by Svalgaard and Schatten (2016) and covers the interval from 1610 to the present day. This is termed the "backbone" reconstruction $R_{BB}$ because the method used is to combine data from various observers into a "backbone" segment and then relate the backbones by regression of annual means. Ostensibly this reduces the number of regressions but, in fact, because regressions (and/or ratios) are often used to extend each backbone and give overlap with the next, this is not actually the case. The authors claim to have avoided daisy-chaining but because there is no method presented to relate early and modern data without relating both to data taken in the interim, this is patently not the case. In constructing $R_{BB}$, the quality of data was assessed by its correlation to the key data sequence on which each backbone is based: however correlation is an inappropriate metric in this context as high correlation can persist even if there are relatively large calibration drifts. Lockwood et al. (2016a) find there is a discontinuity in $R_{BB}$ at the Waldmeier discontinuity implying that the Zürich data, or Zürich procedures (or an over-correction for them), have somehow entered into the construction of $R_{BB}$. A particular concern about the regressions used in constructing $R_{BB}$ is that not only is linearity assumed of the various group number estimates assumed, but also proportionality is assumed.

Lockwood et al. (2016c) point out that there is no advantage to these assumptions, and that they give unreliable regressions (mainly because of non-normal error distributions). The variation of $R_{BB}$ is shown in red in Figure 1(b-i) and its deviation from the median $R_{med}$ in figure 1(b-ii).

## 2.6 The "Corrected" Sunspot Number, $R_C$

Lockwood et al. (2014a) generated a simple "corrected" version of $R_{ISNv1}$ by using a correction for the Waldmeier discontinuity of 11.6% which they derived from two independent statistical techniques using the RGO data. Clette and Lefèvre (2016) present reasons why this correction factor may be too low and this is discussed further in section 3 of the present paper. Lockwood et al. (2014a) also adopted the Leussu (2013) correction to the Wolf data and extended the series back to before 1818 using a daisy-chained regression and appending $1.3R_G$ for 1610-1818, the factor 1.3 being derived by a regression for 1818-1847. The variation of $R_C$ is shown in blue in Figure 1(a-i) and its deviation from the median $R_{med}$ in figure 1(a-ii).

## 2.7 Comparison of Composites

Figure 1 allows comparison of these data series. (Note that group numbers $R_{BB}$ and $R_{UEA}$ have been multiplied by 12.08, as used to generate $R_G$). The $R_C$ variation in figure 1(a) is close to median $R_{med}$ and so the comparisons with $R_{med}$ in the other panels happen to be roughly the same as comparisons with $R_C$.

$R_{BB}$ (figure 1b) is the most radically different of all the composites, giving consistently larger values before 1947 and consistently smaller ones after it. The fractional differences to $R_{med}$ generally increase as one goes back in time. The changes combine to make previous maxima in $R_{BB}$ much more similar to the recent ones so that, whereas all other composites show a fluctuating rise from the Maunder minimum to the recent grand maximum, $R_{BB}$ shows three roughly equal such grand maxima since the Maunder minimum. Furthermore, the variation in $R_{BB}$ has a bistable appearance and so has implications for dynamo models as it suggests that solar activity predominantly exists in either the grand maximum state or the grand minimum state, rather than varying continuously between the two. Lockwood et al. (2016b) show $R_{BB}$ becomes increasingly larger than other solar-terrestrial indicators as one goes back in time;

for example, compared with the observed occurrence of terrestrial aurora at lower magnetic latitudes. This is true at both sunspot minimum and sunspot maximum. Physics-based comparisons with cosmogenic isotopes $^{14}$C, $^{10}$Be and $^{44}$Ti also all show that $R_{BB}$ becomes increasingly too large as one goes back in time (Asvestari et al., 2016). Of these tests, that against $^{44}$Ti abundances is particularly telling because this isotope is measured in meteorites and accumulates slowly as the meteorite is processed on its journey to Earth through the solar system. As a result, the observed $^{44}$Ti is an indicator of the time-integral of solar modulation of the cosmic rays that generate it and so is a sensitive indicator of the long-term changes in solar activity.

The $R_{UEA}$ variation (figure 1c) shows some differences, in both senses, to $R_{med}$. The original group number $R_G$ variation (figure 1d) is consistently lower than $R_{med}$ and is the lowest of all the values in the earlier years. $R_{ISNv1}$ and $R_{ISNv2}$ are both similar to $R_{med}$, the major difference being the effect of allowance (or lack of it) for the Waldmeier discontinuity, with $R_{ISNv1}$ consistently above $R_{med}$ after 1947 (figure 1f) whereas $R_{ISNv2}$ is consistently smaller than $R_{med}$ in this interval (figure 1e).

3. THE WALDMEIER DISCONTINUITY

As discussed above, there is now considerable agreement that the Waldmeier discontinuity is real feature of $R_{ISNv1}$ and that it requires correction in that data series. However there has been debate about how big that correction should be. The smallest correction was derived by Friedli (2016) who finds a correction of just 5%, which applies only to data from 1946-1980. The largest proposed correction was by Svalgaard (2011) who argued that before 1945 sunspot numbers need to be increased by a correction factor of 20%, but it is not clear how this value was arrived at beyond visually inspecting a plot of the temporal variation of the ratio $R_G/R$ (neglecting low values of $R$ below an arbitrarily-chosen threshold), where $R_G$ are the RGO group numbers which were not influenced by Waldmeier's changes to procedures at the Zürich observatory. This assumes that the correction required is purely multiplicative, such that before the discontinuity the corrected value $R' = f_R R$ (and Svalgaard's estimate is $f_R = 1.2$) is required to make the pre-discontinuity values consistent with modern ones (i.e., proportionality is assumed). Because the use of ratios causes an asymmetric distribution of errors and omits sunspot minimum values according to an arbitrarily-chose threshold, Lockwood et al. (2014a) devised two different methods to quantify the discontinuity which give answers that agree very closely, but uncertainties are smaller for the second (so it provides the more stringent test). The first method studies the effect of varying an imposed discontinuity correction factor $f_R$ on the correlation between the sunspot data series tested $R$ and a number of corresponding test sequences (including the RGO $N_G$ values). The second, more stringent, test used fit residuals when $R$ is fitted to the same test data sequences: Lockwood et al. (2014a) then studied the differences between the mean fit residuals before and after the putative Waldmeier discontinuity and quantified the probability of any one correction factor $f_R$ with statistical tests. Because both the sample sizes and the variances are not the same for the two data subsets (before and after the putative discontinuity), these authors used Welch's t-test to evaluate the probability p-values of the difference between the mean fit residuals for before and after the putative discontinuity. This two-sample t-test is a parametric test that compares two independent data samples (Welch, 1947). It was not assumed that the two data samples are from populations with equal variances, so the test statistic under the null hypothesis has an approximate Student's t-distribution with a number of degrees of freedom given by Satterthwaite's approximation (Satterthwaite, 1946). The distributions of residuals were shown to be close to Gaussian and so, as expected, application of non-parametric tests (specifically, the Mann–Whitney U (Wilcoxon) test of the medians and the Kolmogorov–Smirnov test of the overall distributions) gave very similar results. From this quantitative comparison with the RGO $R_G$ data, and assuming proportionality, Lockwood et al. (2014a) derived an 11.6% correction for $R_{ISNv1}$ with an uncertainty range 8.1–14.8% at the 2σ level. The probability of the correction needed being as large as 20%, as advocated by Svalgaard (2011), was found to be $1.6 \times 10^{-5}$.

Clette and Lefèvre (2016) make the valid point that there are other factors which may have influenced the correction factor derived by Lockwood et al (2014a). The first factor is a putative drift in RGO $N_G$ values before 1885 (Cliver and Ling, 2016) which is discussed further in section 5.1 of this paper. This is a relevant factor for the Lockwood et al. (2014a) paper as they used all the RGO data (from 1875), but not for Lockwood et al. (2016a) as they only used data for after 1932. The second potential factor is the precise date of the discontinuity, which is not known because Waldmeier's documentation is not clear when the changes were actually implemented. As discussed by Friedli (2016), the weighting of sunspot groups according to their size might have been implemented (at least partly) by Wolfer and his successors in the beginning of the 20th century. Accordingly, some of the change might be gradual and intermittent. Clette and Lefèvre (2016) make use of means of the ratio $R/R_G$ to define the date of the discontinuity, something that was avoided by Lockwood et al. (2014a) because the error in this ratio tends to infinity when $R_G$ tends to zero and $R_G$ has a minimum in 1944–1945, close to the putative discontinuity and any changes would naturally become more apparent as sunspots began to rise in the next cycle. From the $R/R_G$ ratio, Clette and Lefèvre (2016) place the discontinuity in 1946, although they agree that there is some documentary evidence that at least some of the new procedures that are thought to be the cause of the discontinuity were in use earlier than this date. Clette and Lefèvre (2016) analyse the effects of both the start date of the comparison and the date of the discontinuity assumed for the $R_{ISNv1}$ correction derived by the Lockwood et al (2014a). They reproduced the Lockwood et al. (2014a) values when using the same start and discontinuity dates; however, they found that the correction could be as large as 15.8% for other values of these dates, which is closer to the 18% actually employed in generating $R_{ISNv2}$. Clette and Lefèvre (2016) also report on a study of the inflation caused in a repeat analysis of modern data by adopting Waldmeier's procedures, compared to the results for prior procedures. However, application of such factors assumes knowledge about precisely what procedure was in use and when, and assumes there are no other factors. Also this analysis cannot be used outside the range of the test data as the effect was found to vary non-linearly with the level of solar activity. Hence calibration against other simultaneous data remains the most satisfactory way to evaluate the discontinuity.

Lockwood et al. (2016a) removed any possibility of that early RGO data were having an effect by repeating the study using only data from 1932 onwards (a date chosen to match available ionospheric data) and found a correction factor for $R_{ISNv1}$ of 13.6% using RGO data (and a well-defined value of 12.1% using the ionospheric data). However, this analysis did not take into account the potential effect of the date of the discontinuity.

At this point it should be noted that the analysis of Clette and Lefèvre (2016) applies to sunspot numbers and, as pointed out by Lockwood et al. (2016a), the correction needed for the group numbers generated by Zürich (as part of their derivation of sunspot numbers) will not be the same as that needed for sunspot numbers and that no correction is needed for RGO, or other non-Zürich group numbers. Note that over the 20[th] century there has been a drift in the lifetimes of spot groups, giving an increase in the number of recurrent groups (groups that are sufficiently long-lived to be seen for two or more traversals of the solar disc as seen from Earth) (Henwood et al., 2010). This has the potential to have influenced group numbers derived using different classification schemes in different ways.

Lockwood et al. (2016d) have refined the fit residual comparison procedure yet further. They initially take all available data between 1920 and 1976 (thereby avoiding any effects of both the putative RGO calibration drift and the Locarno error) but omit all data between 1943 and 1949, a six-year interval centred on the optimum date for the discontinuity found by Clette and Lefèvre (2016). Assuming that the bulk of the discontinuity lies within this six-year interval, its precise date is no longer a factor. As also pointed out by Clette and Lefèvre (2016), the longer the intervals used in the test, the greater is the chance that other errors and discontinuities in either the test or the tested data become a factor. On the other hand, if the intervals used are too short, then the uncertainties inherent in the method (indeed in all such comparison methods) get larger because of the geophysical noise variability in the data series. To find the optimum interval, Lockwood et al. (2016d) used a basket of test data series and varied the duration of the "before" and "after" intervals until the net uncertainty was minimised. They also used $2^{nd}$-order polynomial fits so that assumptions of both proportionality and linearity were avoided. The analysis was repeated with $3^{nd}$-order polynomial fits but some of the fit-residual Q-Q plots began to show non-Gaussian distribution tails and so these fits were not used further. To reduce the number of variables in this parametric study, Lockwood et al. (2016d) required the "before" interval and the "after" intervals be of the same duration. Minimum uncertainty (i.e. optimum agreement between the results for the various test data) was obtained using "before" and "after" intervals that were 11 years in duration and hence the "before" data were from 1932−1943 and the "after" date from 1949−1960. In addition, Lockwood et al. (2016d) did not assume that the correction needed is just a multiplicative factor or even linear but allowed for both a zero-level offset δ and non-linearity in $R$, as well as a sensitivity change (hence they evaluate the corrected series $R' = f_R R^n + \delta$ for "before" interval that is consistent with the "after" interval). Lockwood et al. (2016d) used a wide variety of test data in addition to the RGO group number $[N_G]_{RGO}$, namely: total sunspot area $A_G$ from the RGO dataset; the CaK index from the Mount Wilson spectroheliograms in Ca II K ion line; the sunspot group number from the Mount Wilson sunspot drawings, $[N_G]_{MWO}$; and the ionospheric F2 region critical frequencies measured by the Slough ionosonde, foF2. They tested all six of the sunspot series discussed in the introduction using these five test series. By multiplying the probability distribution functions for the five tests together, Lockwood et al. (2016d) obtain the optimum correction for each sunspot data series for around the Waldmeier discontinuity, a procedure that has the advantage of weighting the overall estimate according to how well-constrained each individual value is. Note that for all tested series, the narrowest pdf (and hence the most well-defined value, thereby automatically gaining most weighting) was the RGO group numbers but the optimum values for the other test data series always agreed to within the $\pm 2\sigma$ uncertainty band for the RGO group number data. Lockwood et al. (2016d) estimate the correction factors needed for the six composites discussed here.

For $R_{ISNv2}$ and $R_{UEA}$ (which equals $R_G/12.08$ over the interval studied) it was found that the exponent $n$ was near unity and the offset $\delta$ was very small. Thus the corrections required were approximately linear. However, this was not found to be true for $R_{ISNv1}$, $R_{BB}$ and $R_C$. To quantify the magnitude of the discontinuity in each tested data sequence, Lockwood et al. (2016d) evaluated the percentage change over the "before" interval 1932–1943 (approximately solar cycle 17). Note however, in the case of $R_{ISNv1}$, $R_{BB}$ and $R_C$, the non-linearity of the correction required means that this percentage change cannot simply be applied to all the prior solar cycles.

For ($R_G/12.08$) and $R_{UEA}$, Lockwood et al. (2016d) found the net correction required to the "before" interval is $+0.005\% \pm 0.05\%$. This is no more than a test of the procedure as both ($R_G/12.08$) and $R_{UEA}$ are the RGO group number data for both the before and after intervals, which is the dominant test series and hence the correction factor should indeed be zero. The uncertainty arises from the effect of the other test datasets used, in addition to RGO group numbers, and the low value of this uncertainty stresses the level of agreement between the test datasets.

For $R_{ISNv1}$, Lockwood et al. (2016d) found the net correction required to the "before" interval is $+12.3\% \pm 3.4\%$. This is larger than the 11.9% used by Lockwood et al. (2014a) but smaller than the 15.8% derived by Clette and Lefèvre (2016); however, it almost agrees with both to within the $2\sigma$ uncertainties. The study also finds that the changes introduce by Waldmeier had a somewhat non-linear effect as the optimum exponent $n$ is 1.088.

The above correction to $R_{ISNv1}$ is significantly smaller than the 18% used in the derivation of $R_{ISNv2}$. This is consistent with the correction for $R_{ISNv2}$ in the "before" interval found by Lockwood et al. (2016d) which is −3.8% ± 2.9%. This is not quite zero, to within the derived 2σ uncertainties. Hence the best estimate from this study is that $R_{ISNv2}$ is based on a slight over-correction for the Waldmeier discontinuity. Note however, that the non-linearity of the discontinuity in $R_{ISNv1}$ (i.e. the fact that different group number levels are affected differently, making $n$ different from unity) has been successfully removed in $R_{ISNv2}$ as the optimum $n$ in this case was found to be 0.997.

For $R_C$, the correction for the "before" interval is +0.4% ± 3.0%. Note, however, that the non-linearity inherent in $R_{ISNv1}$ was found to persist ($n = 1.095$) and the simple corrections used in $R_C$ means that it carries forward other errors in $R_{ISNv1}$, such as the Locarno calibration drift. Hence, although it matches cycle 17 slightly better than does $R_{ISNv2}$, in several ways it is a less satisfactory correction.

For $R_{BB}$, Lockwood et al. (2016d) found the net correction required to the "before" interval is −5.7% ± 2.2%, i.e. there is, effectively, an over-correction for the Waldmeier discontinuity and by more than that for $R_{ISNv2}$. Furthermore, the non-linear behaviour has not been removed ($n = 1.093$).

## 4. COMPARISON WITH OPEN SOLAR FLUX RECONSTRUCTIONS

Observations of geomagnetic activity were first made in 1722 by George Graham in London. In 1798 Alexander von Humboldt made observations from a number of locations, work that sparked the interest of his friend, Carl Friedrich Gauss, who developed the first reliable and stable magnetometer and so established the first magnetic observatory in Göttingen in 1832. Although we have fragments of data from before 1845, Lockwood et al. (2013a; b; 2014c; d) considered that only after this date can we compile (for the time being at least) homogeneous and well-calibrated geomagnetic data sequences. This is true for both hourly means of the field components and for "range" indices, based on the range of variation of components within 3-hour intervals.

The big advantage of geomagnetic observations is that they are instrumental measurements that, unlike sunspot numbers and sunspot group numbers, involve no subjective decisions by the observer. Because they are closely related to sunspot numbers they offer a potential way to evaluate and check sunspot number records (e.g., Svalgaard and Cliver, 2007). The method first used by Wolf was to look at the quiet day diurnal variation in geomagnetic activity, now understood to be due to thermally driven thermospheric winds but varying mainly with the ionospheric conductivity, and hence the ionising EUV flux from the Sun (Brekke et al., 1970). As the EUV flux has a close correlation with sunspot numbers, this could provide a means of calibration of sunspot numbers. However, the driving thermospheric winds also vary with sunspot numbers, but with a different dependence to the conductivities (e.g., Aruliah et al., 1996) and also show long-term trends that are not of solar origin (Bremer et al., 1997). In addition, the secular variation in the geomagnetic field influences ionospheric conductivities and hence the quiet-day magnetic variations (Cnossen and Richmond, 2013; de Haro Barbas, 2013). These factors give variability in the relationship between sunspots and the quiet-day geomagnetic variation that is unknown, which, although small, is still sufficient to make this calibration unreliable. For example, Svalgaard and Cliver (2007) find that sunspot numbers and the quiet day geomagnetic variation have a correlation coefficient of $r = 0.985$ with the international sunspot number $R_{ISNv1}$ which leaves a 3% variation that is unexplained ($r^2 = 0.97$) –in addition $R_{ISNv1}$ is now known to contain errors. Tests show that even this very high $r$ could disguise a drift in $R_{ISNv1}$ of up to 0.1 yr$^{-1}$ which would amount to 50% of the mean value over the interval between 1750 and the present. Hence correlation is not an appropriate metric for assessing the potential of a proxy dataset to provide calibration.

An alternative opportunity to use geomagnetic data in this context arises from the facts that the hourly mean data depend primarily on the near-Earth interplanetary magnetic field (IMF) and the range indices depend on both the IMF and the solar wind speed (see discussion and explanation in Lockwood, 2013). This allows reconstruction of the "open solar flux" (OSF, also called the "heliospheric source flux": here we used the signed OSF, denoted by $F_S$) from combinations of hourly mean and range geomagnetic data (Lockwood et al., 2014d). OSF provides a good test for sunspot numbers because it is, like sunspot number, a global indicator of solar magnetism, rather than a local heliospheric parameter such as the near-Earth solar wind speed and IMF (although, as discussed by Owens et al. (2016), there is still a close relationship between sunspot number and near-Earth interplanetary magnetic field ). In addition, the variation of $F_S$ is determined by a continuity equation in which the source term has been expressed in terms of sunspot numbers by Solanki et al. (2000) who used it to model the $F_S$ variation reconstructed from the aa geomagnetic index by Lockwood et al. (1999). The model has evolved subsequently for various applications with refinements to both the production and loss rate formulations used (Lockwood, 2003; Owens and Crooker, 2006, 2007; Vieira and Solanki, 2010; Schwadron et al., 2010; Owens and Lockwood, 2012; Goelzer et al., 2013; Lockwood and Owens, 2014). A development used here are cycle-dependent OSF loss rates: from theory and observations of coronal inflows (Sheeley et al., 2001), loss rates that depend on the tilt of the heliospheric current sheet were predicted by Owens et al. (2011). Owens and Lockwood (2012) showed that the implied variation of the OSF loss rate with the phase of the solar cycle arose naturally for the suggested dependence of the OSF source on sunspot numbers and the reconstructions of OSF from geomagnetic activity data.

In parallel to this modelling development, reconstructions of OSF from geomagnetic activity indices have been refined (see review by Lockwood, 2013). The most sophisticated and robust is that by Lockwood et al. (2014d) who used four pairings of geomagnetic indices and Monte Carlo techniques to estimate all uncertainties and combine the results from the four pairings. Recent work reveals the great extent to which this gives robustness against possible calibration errors in any one geomagnetic data series (Lockwood et al., 2016e). This OSF reconstruction allows for the effect of the solar wind speed on the Parker spiral garden hose angle, and for the effect of "folded flux" that threads the heliocentric sphere of radius 1AU more than once, thereby making the flux through that surface greater than the OSF by an amount termed the "excess flux" (Lockwood and Owens, 2009).

The OSF reconstruction from geomagnetic activity data is also completely independent of the sunspot data. There is one solar cycle for which this statement needs some clarification. Lockwood et al (2013a) used the early Helsinki geomagnetic data to extend the reconstructions back to 1845 and Svalgaard (2014) used sunspot numbers to identify a problem with the calibration of the Helsinki data in the years 1866−1874.5 (much of solar cycle 13). Lockwood et al. (2014c) re-evaluated the Helsinki data using simultaneous data from the nearby St-Petersburg magnetometer and a study of the modern-day data from the nearby Nurmijarvi station. The results confirm the conclusion of Svalgaard (2014) but it is important to stress that the correction of the Helsinki data for solar cycle 11 made by Lockwood et al (2014c), and subsequently used by Lockwood et al (2014d), was based entirely on magnetometer data and did not use sunspot numbers, thereby maintaining the independence of the two datasets. The geomagnetic OSF reconstruction provides a better test of sunspot numbers than the quiet day geomagnetic variation because the uncertainties in the long-term drift in the relationship between the two are understood and have been quantified.

The formulation of the OSF model used here was a follows. As used by Owens and Crooker (2006), the OSF source term, $S$, is assumed to follow the CME rate, on average. The best fit between observed CME rate (e.g., Yashiro et al., 2004) and $R$ gives $S = \phi$ (0.234 $R^{0.540}$ – 0.00153) Wb per Carrington rotation, where $\phi = 10^{12}$ Wb is the average closed flux carried by a CME (Lynch et al., 2005; Owens, 2008). For each sunspot record, the loss term, $L$, is computed by subtracting $S$ from the rate of change of geomagnetic OSF estimates over 1845-present. For all sunspot records, the fractional $L$ shows a strong solar cycle variation, but remarkably little cycle-to-cycle variation (Owens and Lockwood, 2012), in close agreement with the heliospheric current sheet (HCS) tilt variation, as expected from theory (Sheeley et al., 2001; Owens et al., 2011). From the $L$ time series, we calculate the average fractional $L$ as a function of solar cycle phase, which is used with $S$ to compute sunspot-based estimates of OSF. The scatter between the sunspot- and geomagnetic-based estimates of OSF over 1845-present are used to quantify the uncertainty in $R$-based estimate (i.e., the geomagnetic OSF estimate is assumed to represent the ground truth).

Figure 3 shows the OSF model results for the sunspot number and sunspot group number sequences shown in figure 1, using the same colours. In each panel, the black line is the geomagnetic reconstruction of Lockwood et al. (2014a) with the ±1σ uncertainty band shown in grey. The coloured line is the best fit for the sunspot number/sunspot group number used and the lightly coloured area is the ±1σ uncertainty for that fit. The darker coloured region is where the two uncertainty bands overlap. It can be seen that the model captures the main features (the decadal-scale solar cycle variations and centennial-scale drifts) very well for all of the input sunspot data sequences. This shows that the model is not relying on a feature of any one of the sunspot number sequences. The one exception to these statements is solar cycle 20, for which all of the sunspot sequences fail to reproduce the flat-topped appearance of the OSF variation. It is tempting to ascribe this to an error in the geomagnetic OSF reconstruction, however, this is not the case as solar cycle 20 is covered by in-situ interplanetary observations and these match the geomagnetic reconstruction very well (Lockwood et al., 2014a). A possible explanation may lie in the effect of the sunspot tilt angle which quantifies the difference in latitude of the two footpoints of the associated bipolar magnetic region field loops. This influences the speed with which they separate under differential rotation and hence the upward evolution of the loop through the corona (MacKay et al., 2002, MacKay and Lockwood, 2002). Using a flux transport model with solar-cycle averages of observed sunspot tilt angles, Cameron et al. (2010) are able to reproduce the OSF in cycle 20 very well and average tilt angles are considerably lower during the exceptionally strong preceding cycle (number 19) than for all other cycles. Because sunspot tilt angle data are only available continuously after 1918, their potential effects on the source rate $S$ are not allowed for in the model used here.

Table 1 gives the fit parameters in each case: $r$ is the correlation coefficient, $S_r$ the significance of $r$ (allowing for the persistence in the data and comparing against the AR1 noise model), $\Delta$ is the r.m.s. difference between the reconstructed and fitted OSF values, $\Delta_P$ is the r.m.s. difference between the reconstructed and fitted OSF values for three year intervals around the solar-cycle maxima in OSF (peaks), $\Delta_T$ is the r.m.s. difference between the reconstructed and fitted OSF values for three year intervals around the solar-cycle minima in OSF (troughs). There are no statistically significant differences between these fits. The best fit, according to several metrics, is for $R_{ISNv2}$ which shows an improvement over the fit for $R_{ISNv1}$ in all metrics. The group numbers do not fare quite as well, which is to be expected as sunspot group number is unlikely to be as good a proxy of total solar magnetic flux emergence through the photosphere and coronal source surface as sunspot numbers. Of these, the fits for $R_{UEA}$ and $R_{BB}$ are very slightly better than that $R_G$. However none of these differences are significant at even the 1σ level. Looking closely at figure 3, some qualitative differences between the fits do become apparent.

Figure 3(a) shows the results for $R_C$ (in blue). The modelled and reconstructed OSF sequences are very similar except for cycle 9 (the first one in the sequence) when the value derived from $R_C$ is too low. As discussed below, this occurs for several of the sunspot data sequences. A major success is that in addition to the long-term variation, this fit matches the solar cycle amplitudes, reaching down to the minima and up to the maxima. The is no change detected across Waldmeier discontinuity which one might expect to see if the correction used was grossly in error.

Figure 3(b) shows the results for $R_{BB}$ (in red). Again, this yields a larger OSF in cycle 9 but elsewhere the fits are not as close as for $R_C$ in that $R_{BB}$ shows a tendency to underestimate solar cycle amplitudes and there is a strong suggestion of over-correction for the Waldmeier discontinuity with peak values being subsequently too low.

Figure 3(c) shows the results for $R_{UEA}$ (in orange). Unlike $R_C$ and $R_{BB}$, this reproduces the OSF variation in solar cycle 9 well, however it does underestimate them in cycles 10 and 11 and the amplitudes of cycles 14, 15 and 16 are very slightly overestimated. Figure 2(d) shows the results for $R_G$ (in green), which are very similar to those for $R_{UEA}$.

Figure 3(e) shows the results for $R_{ISNv2}$ (in mauve). There may be a slight tendency to underestimate peak values and solar cycle amplitudes after the Waldmeier discontinuity, but it is not as marked as for $R_{BB}$.

Figure 3(f) shows the results for $R_{ISNv1}$ (in brown). There is a marked tendency to overestimate cycle peaks after 1947, consistent with the Waldmeier discontinuity. Note that the tendency for over-estimation of modern cycles using $R_{ISNv1}$ is as great as the tendency for under-estimation in the same cycles for $R_{BB}$.

## 5. OBSERVER SCALING FACTORS INHERENT IN RECONSTRUCTIONS

The $k'$ factors at a given level of solar activity used in generating a group numbers are usually assigned to an observer and assumed to stay constant over the duration of his/her observing lifetime. However, a number of factors may vary on a range of timescales for a given observer: these include atmospheric conditions, local site conditions (for example via stray light), equipment used, the algorithms, metrics and procedures that the observer adopted to help make the subjective decision as to what constitutes a sunspot group and even his/her eyesight. These factors can introduce long-term drift as well as year-to-year variability in the data from each observer. We can assess the drifts and variability for each observer that are required by each of the reconstructed group number composites. We do this by studying the variations of annual observer $k'$ factors, $k_a' = R_g/<N_G>$, inherent in a generic sunspot group number reconstruction $R_g$ and where $<N_G>$ is the annual mean of the raw sunspot group number count by the observer in question. In this section we consider the implications of both $R_{BB}$ and $R_{UEA}$ for observers active in the second half of the 20[th] century.

Figure 4 plots annual means of the group numbers $R_G/12.08$ (in green – note the normalising factor in equation (2) has been cancelled), $R_{BB}$ (in red), and $R_{UEA}$ (in orange). The black line is the "Schwabe backbone", $R_{BBS}$, generated by Svalgaard and Schatten (2016) which they multiply by 1.48 to obtain $R_{BB}$, that being the factor that they derive from linear regression (assuming proportionality) between their Schwabe and Wolfer backbones over 1861−1883. It can be seen that there is a significant difference between $R_{BB}$ and $R_{UEA}$ before 1885 and that this is mainly explained by this calibration of the two backbones because $R_{BBS} = R_{BB}/1.48$ (in black) is very similar indeed to $R_{UEA}$ (in orange). An additional factor is a putative drift in the RGO group number data calibration which has been proposed by Cliver and Ling (2016) to be present. The factors combine to make $R_{BB}$ considerably larger than both $R_{UEA}$ and $R_G/12.08$ and they are investigated in this section.

**5.1 The drift in early RGO data**

The top panels of figure 5 show the annual $k_a'$ factors for various observers that are inherent in (a) $R_{BB}$ and (b) $R_{UEA}$. Ideally, each observer would not vary in data quality and give $k_a'$ points that lie along horizontal lines (i.e. $k_a'$ is a constant, $k'$, at all times). Noise (interannual variability) can be averaged out by taking a mean for that observer over several years (i.e. $k' = <k_a'>$), but trends in $k_a'$ mean that either the observer's data quality changed over time or that the reconstructed group number used to compute $k_a'$ is in error. This is significant because if several observers' $k_a'$ values show the same trend, the common denominator is the reconstructed group number which would then be inferred to be in error. Figure 5 shows that both $R_{BB}$ and $R_{UEA}$ give observers' $k_a'$ values that reveal, in general, both year-to-year variability and longer-term drifts.

At sunspot minima (the joins between grey and white vertical bands in figure 5), large values of $k_a'$ are often seen. This means that the reconstructed composite is not reaching down to as low minimum values as the observations and is a consequence of the asymmetric uncertainties in taking ratios which become large at sunspot minimum. This occurs for $R_{UEA}$ around 1890 (the minimum between solar cycles 12 and 13) and for both $R_{BB}$ and $R_{UEA}$ around 1879 (the minimum between solar cycles 11 and 12). This does not mean the reconstructions are incorrect at these minima, but a low acuity observer could be observing proportionally fewer spot groups at sunspot minimum, as discussed by Lockwood et al.

(2016c). Indeed the realisation by Wolf (1873) that $k$ and $k'$ factors depend on the level of solar activity tell us that we should, in general, expect this behaviour.

Looking at the averages of $k_a'$ for either reconstruction, it is clear that the observers have considerably different $k'$ factors. We here normalise the $k_a'$ values by dividing by the mean for a reference period. To avoid the effect of the large asymmetric errors at sunspot minimum where here use the interval 1883–1888 for that reference period, which spans the approximate date of 1885 for the putative discontinuity in the RGO data, as defined by Cliver and Ling (2016). The black dots show the results for all data excepting the RGO data, the yellow dots show the RGO data. The red histogram gives the mean for all the black dots (i.e., excluding the RGO data). Using $R_{BB}$, the calibration drift noted by Cliver and Ling (2016) is seen as the increasing difference between the red histogram and the yellow dots as one goes back in time. Both the red histogram and the yellow dots show greater variability for $R_{UEA}$ than for $R_{BB}$, but no great importance should be placed on this as it relates to very small differences at sunspot minimum. However, it is significant that the RGO data and the mean of the other data have very similar variations after about 1885, except that in both the lower panels of figure 5 we seen that the RGO data are a bit lower than the mean of the observers data for 1892–1895 (inclusive). Cliver and Ling (2016) state that the onset of the discontinuity in the RGO data (as we go back in time) is about 1885 but figure 5 shows that RGO values remain close to the mean of the available observers for 1882-1885 and only are too small for 1875–1881. Even then, the 1881 value is not significantly low (it is within the spread of other observers) and the 1879 and 1880 values are at sunspot minimum and so are exaggerated by taking ratios. Hence we agree with Cliver and Ling's (2016) conclusion that the earliest RGO data are too low; however, the problem is largely confined to the first three years of the data series (1875–1877, inclusive) in the declining phase of solar cycle 11. We also note that a second period, not mentioned by Cliver and Ling, when the RGO values are systematically too low compared to other observers exists in the years 1892–1895. Looking at the mean values given by the red histograms, for $R_{BB}$ they increase slightly but systematically with decreasing time from unity in 1882 to 1.1 in 1874. Thus although the drift in RGO calibration appears to be real, it is exaggerated in comparisons with $R_{BB}$ by a ≈10% drift in $R_{BB}$, relative to the mean of the basket of available observers. Looking in the green, red and blue points in the top panel of figure 5(a) at this time we can see that this drift is also revealed by comparison with the data from Wolfer, Wolf and Schmidt (respectively) but not in the data from Moncalieri and Tacchini (black and light grey dots, respectively, which remain at a near constant $k_a'$) and the Spörer data (in orange, for which $k_a'$ actually varies in the opposite sense). In the corresponding figure for $R_{UEA}$ (figure 5b) all these data series remain more constant and the 1875−1878 values are within the range of variations seen in previous years and this is even true for the Spörer data, except for the year 1876. The increase in $R_{BB}$, relative to the average of a basket of observers, in the first few years of the RGO data is critical to the $R_{BB}$ data series because of the daisy-chaining method used: before 1883 is the overlap period used to calibrate the Schwabe and Wolfer backbones, which means this drift affects all previous data. Note that no RGO for before 1900 were used in the construction of $R_{UEA}$.

## 5.2 Intercalibration of the data of Schwabe and Wolfer

A key intercalibration for daisy-chained composites (i.e., all but $R_{UEA}$) within the interval studied here is that between the data of Schwabe and Wolfer, as these data form key parts of all constructions of a centennial-scale sunspot activity index. The Schwabe data cover 1826 to 1867 whereas the Wolfer data cover 1878 to 1928. In the construction of $R_{BB}$, the Schwabe data are extended to later times, and the Wolfer data extended to earlier times, using data from other observers to generate the "Schwabe backbone" and "Wolfer backbone" respectively. Note that the same data are used to extend both backbones. The Schwabe backbone is then re-calibrated to the Wolfer backbone using linear regression (also assuming proportionality) over the interval 1861−1883.

Part (b) of figure 4 shows the interval 1874−1920 in more detail. This includes the interval 1874−1885 for which the RGO data calibration has been questioned (Cliver and Ling, 2016) and which, as discussed above, has an effect on the calibration of all data for earlier times if daisy-chaining is employed. Before 1900, the Usoskin et al (2016) reconstruction $R_{UEA}$ does not use the RGO data and for the interval over which the RGO calibration has been questioned, $R_{UEA}$ includes the data recorded by Wolfer (1876 – 1928), Winkler (1882 – 1910), Tacchini (1871 – 1900), Leppig (1867 – 1881), Spörer (1861 – 1893), Weber (1859 – 1883) and Wolf (1848 – 1893). It is important to remember that all of these data have been calibrated, independently of each other, using the active-day fraction method and comparing against RGO data for after 1900. Figure 4(b) shows that despite adding all these data, for 1874–1900, $R_G$ (i.e. the RGO group number data, green line) and $R_{UEA}$ (orange line) remain very similar indeed.

Figure 6 is in the same format as figure 5, but studies the join between the Schwabe and the Wolfer data. The observers shown are all those used in the construction of $R_{BB}$ that produced data that spanned 1872, which is in the centre of the gap between the Schwabe and Wolfer datasets. Hence these are the observations (and the only observations) used to extend to the two backbones and hence intercalibrate the Schwabe and Wolfer data in the construction of $R_{BB}$. Those observers were: Spörer (shown in orange); Wolf (using the small telescope, shown in red); Schmidt (blue); Tacchini (grey); Leppig (mauve); Weber (pink); Howlet (cyan) and Meyer (brown). The Schwabe data are shown in yellow and the Wolfer data in green. In order to visually highlight the variation of the $k_a'$ factors for each observer, a second order polynomial was fitted for each observer to help identify trends whilst supressing the year-to-year variability.

Considering $R_{BB}$, the top panel of figure 6(a) shows that $R_{BB}$ predicts that the $k_a'$ factor for Wolf's small telescope data drifted down with time very slightly throughout the interval that he took such measurements (red line); this implies he as measurements got slightly more accurate over time. This is somewhat surprising as $k$ and $k'$ factors for Wolf have generally been thought to increase due to his deteriorating eyesight, which is also found in the study by Friedli (2016) (see his figure 10). For Spörer (orange) and Schmidt (blue) the $k_a'$ factor initially fell but then rose again (implying these observers initially grew in acuity but later grew less able to detect spot groups): for the intercalibration interval of 1861−1883, the $k_a'$ values for Spörer are almost constant whereas they rise consistently with time for Schmidt; for Tacchini (grey) the $k_a'$ are constant but these data only cover the second half of the calibration interval; for Leppig (mauve) $k_a'$ fell rapidly with time but these data only cover the middle of the calibration interval; for Weber (pink) it was initially constant but then rose rapidly; for Howlet (cyan) $k_a'$ initially fell very rapidly but then levelled off ; and for Meyer (brown), $k_a'$ fell rapidly but these data only cover the first half calibration interval. Thus the results of intercalibration of Wolfer and Schwabe will depend critically on the observer used to pass on the calibration. The data of Spörer and Schmidt argue that the data of Schwabe (up to 1867) is correctly joined to that of Wolfer (after 1878) in $R_{BB}$, whereas the data of Wolf argue that in $R_{BB}$ the Schwabe data have been inflated somewhat and the data of Leppig,

Howlet, and Meyer argue that it is inflated by a large factor. On the other hand, the data of Weber argues that it has not been inflated enough.

To take an average of these results, the bottom panel of figure 6(a) shows the average variation as a red histogram, generated in the same way as for the bottom panels of figure 5. The reference period used to normalise the $k_a'$ values is $1868-1876$ which avoids sunspot minimum years for the reasons described above.

The same procedure was applied to $R_{UEA}$, and the results are shown in figure 6(b). The $k_a'$ values are all smaller and so $R_{UEA}$ is calling for less adjustment of the observers' raw data than does $R_{BB}$. The pattern of drifts is similar (because $R_{UEA}$ and $R_{BB}$ are so highly correlated). We here highlight not so much the average result but the diversity of the results depending on what weight one gives to the different observers. The main point we are making is that daisy-chaining by regression is an inherently unsatisfactory approach and is greatly influenced by a number of subjective decisions about which data to use and over which intervals. This confirms the concerns listed in the introduction. We also note that this calibration interval is actually relatively well populated with data compared to earlier ones.

That having been said, figure 6(a) does provide some evidence that $R_{BB}$ has been inflated going backward in time across this join, as Lockwood et al. (2016c) predicted it would be by the use of non-robust regression procedures and, in particular the assumption that the data series are proportional. The bottom panel of figure 6(a) shows that the mean of the basket of available observations (the red histogram) displays a rise across the calibration interval. (The horizontal blue line is unity). On the other hand, although $R_{UEA}$ does show the large deviations that are to be expected at solar minimum, it gives normalised $k_a'$ values that return to unity, showing no drift across the inter-calibration interval. To illustrate the effects of this, the Schwabe and Wolfer data have been matched to the $k_a'$ for $R_{BB}$ by normalising such that the means are the same over their period of overlap with the red histogram. The results are shown by the yellow and green dots in the bottom panel of figure 6(a). It can be see a clear jump is introduced by the intercalibration and that this is of order 20%. This would argue that the factor of 1.48 used by Svalgaard and Schatten (2016) in constructing $R_{BB}$ is 20% too large and should be nearer 1.2: however, this value is only indicative and we do not advocate its use because the individual observers give widely differing values: the more important point is that this value can be altered by any one several subjective decisions about which data to use and how to carry out the intercalibration, making the intercalibration unreliable.

## 6. DISCUSSION & CONCLUSIONS

We find that proportionality of annual means of the results of different sunspot observers is generally invalid and that assuming it causes considerable errors in the long-term variations of sunspot data composites. This is a particular problem when daisy chaining of calibrations is used as errors accumulate over the interval.

Our analysis of the join between the Schwabe and Wolfer data sunspot series shows the uncertainties in daisy-chaining calibrations are large and demonstrates how much the answer depends upon which data are used to make such a join. This example, which is well-populated with data compared to earlier backbone joins in $R_{BB}$, demonstrates just how unreliable daisy-chaining of calibrations is. The concern highlighted here relates to the quality and variability of the data used to pass the calibration from one data series to the next. In addition to this, the analysis of Lockwood et al (2016c) shows that great care needs to be taken to ensure that linear regressions are not giving misleading results because the data are violating the assumptions of the techniques used. Lastly, Lockwood et al (2016c) and Usoskin et al. (2016) also show that the practice of assuming proportionality, and sometimes even linearity, between data series (and hence using ratios of sunspot numbers) is also a cause of serious error.

Opportunities for quality control of sunspot composites are very limited because if data are good enough to form a test, the scarcity of reliable data means that we always would want to include them in the composite. Thus we have to use quality assurance which means we always rigorously stick to best practice and expunge all broad-brush dismissals as "small" of the effect of any one assumption or approximation. Errors in any intercalibration (whether they are inside a data "backbone" or between them) will compound over time if daisy chaining is used. For this reason we strongly recommend both daisy-chaining and regression procedures are avoided and that the long-term variations in any data composite compiled using either technique, or worse still both, should not be trusted. The only published composite that uses neither daisy-chaining nor regression, nor does it assume proportionality (or even linearity) between the results of different observers, is $R_{UEA}$ by Usoskin et al. (2016).

However, we note that the result of another daisy-chain-free method by Friedli (2016), which is yet to be published, agrees very well with $R_{UEA}$. This is not to say that the $R_{UEA}$ reconstruction has been refined to its optimum possible form. For example $R_{UEA}$, like other composites, currently assumes that observers maintained a constant $k'$ factor (at a given $R$) over the period for which they made observations. This assumption has to be made for daisy-chaining but does not have to be made when every data segment is calibrated by reference to a single standard dataset and interval, as is the case for $R_{UEA}$. However, if the observers' data are sub-divided into too many short segments, the calibration of each will became poorer because the statistics are poorer. We recommend that, as in the analysis of Lockwood et al. (2016d), the duration of the intervals used could be iterated until the optimum compromise is achieved.

Lastly, we need to dispel some misconceptions about any relationship of all the sunspot number reconstructions discussed here to terrestrial climate change. This stems from a press release issued by the International Astronomical Union (IAU) when the backbone group sunspot number was first published (IAU, 2015). This suggested that the lack of gradual change in solar activity in the backbone reconstruction argued against long-term solar change as a major cause of terrestrial climate change: a somewhat bizarre conclusion because there are many, and very much more compelling, scientific arguments behind the scientific consensus that only a minor part of current climate change can be attributed to solar change (IPCC, 2013). We stress that our concerns about the backbone reconstruction are because it uses unsound procedures and assumptions in its construction, that it fails to match other solar data series or terrestrial indicators of solar activity, that it requires unlikely drifts in the average of the calibration $k'$ factors for historic observers and that it does not agree with the statistics of observers' active day fractions. The evidence is that the issues discussed in the present paper do not impinge in any way upon humankind's understanding of terrestrial climate change. We refer the reader to reviews of the effects of solar activity on global and regional climates by Gray et al. (2010) and Lockwood (2012) and the contribution of Working Group 1 to the Fifth Assessment Report of the Intergovernmental Panel on Climate Change (IPCC, 2013). There is growing evidence for, and understanding of, some solar-induced regional climate changes (which almost completely cancel on a global scale), induced by jet stream modulation in winter by changes to stratospheric heating gradients (Lockwood, 2012; Ineson et al., 2015; Maycock et al., 2015), but many studies have found solar effects on global mean temperature are found to be very small (e.g. Jones et al., 2012) and in this context, the difference between the backbone and any other sunspot reconstruction is minimal and of little consequence (Kopp et al., 2016).

**Acknowledgements**.  The authors are grateful to a great many scientists for provision of the sunspot group count data used here, including: David Hathaway, Rainer Arlt, Leif Svalgaard, the staff of the Mount Wilson Observatory, the Solar Observatory of the National Astronomical Observatory of Japan, the Solar Physics Groups at NASA's Marshall Space Flight Center and Mullard Space Science Laboratory and the libraries of the Royal Society of London and the Royal Astronomical Society, London.  The work of M. Lockwood, M.J. Owens and L.A. Barnard at Reading was funded by STFC consolidated grant number ST/M000885/1 and that of I.G. Usoskin was done under the framework of the ReSoLVE Center of Excellence (Academy of Finland, project 272157).

REFERENCES

Aparicio, A.J.P., Vaquero, J.M., Gallego, M.C. 2012, J. Space Weather Space Clim. 2, UNSP A12.  doi:10.1051/swsc/2012012.

Aruliah, A.L., Farmer, A.D., Rees, D., & Brändström, U. 1996, J. Geophys. Res., 101(A7), 15701, doi:10.1029/96JA00360.

Asvestari, E., Usoskin, I.G., Kovaltsov, G.A., Owens, M.J. and Krivova, N.A. 2016, Solar Phys., submitted

Brekke, A., Doupnik, J. R., & Banks P.M. 1974, J. Geophys. Res., 79, 3773, doi: 10.1029/JA079i025p03773

Bremer, J., Schminder, R., Greisiger, K.M.,  Hoffmann, P.,  Kürschner, D., & Singer, W. 1997, J. Atmos. Sol.-Terr. Phys., 59 (5), 497, doi: 10.1016/S1364-6826(96)00032-6

Çakmak, H. 2014, Exp. Astron. 37, 539. doi: 10.1007/s10686-014-9381-6

Cameron, R.H., Jiang, J., Schmitt, D., & Schüssler, M. 2010,  Astrophys. J., 719:264–270

Clette, F. & Lefèvre, L. 2016, Solar Phys., in press. arXiv:1510.06928

Clette, F., Svalgaard, L., Vaquero, J.M., Cliver, E.W. 2015, Revisiting the sunspot number, *in The Solar Activity Cycle*, eds. A. Balogh,  H. Hudson, K. Petrovay and R. von Steiger, 35, Springer, New York. doi: 10.1007/978-1-4939-2584-1_3

Cliver, E.W., & Ling, A. 2016, Solar Phys. pre-published on-line, doi: 10.1007/s11207-015-0841-6

Cliver, E.W., Clette, F., Svalgaard, L. 2013, Cent. Eur. Astrophys. Bull. **37**, 401. ISSN 1845–8319

Cnossen, I., & Richmond, A.D. 2013, J. Geophys. Res., 118, 849, doi: 10.1029/2012JA018447

de Haro Barbas, B.F., Elias, A.G., Cnossen, I., & Zossi de Artigas, M. 2013, J. Geophys. Res. Space Physics, 118, 3712, doi:10.1002/jgra.50352.

Friedli,T.K. 2016, *"The construction of the Wolf Series from 1749 to 1980"*, Solar Phys., submitted.

Goelzer, M.L., Smith, C.W., & Schwadron, N.A. 2013, J. Geophys. Res. Space Physics, 118, 7525, doi: 10.1002/2013JA019404

Gray, L.J., Beer, J., Geller, M., et al. 2010, Rev. Geophys., 48, RG4001, doi:10.1029/2009RG000282, 2010

Hathaway, D. 2015, Living Rev. Solar Phys. 12, 4, doi: 10.1007/lrsp-2015-4, http://www.livingreviews.org/lrsp-2015-4

Henwood, R., Chapman, S.C., & Willis, D.M., 2010, Solar Phys., 262 (2) 299-313, doi: 10.1007/s11207-009-9419-5

Hockey, T. 2014, Biographical Encyclopedia of Astronomers, 2278-2278, doi: 10.1007/978-1-4419-9917-7_1436

Hoyt, D.V., Schatten, K.H. and Nesme-Ribes, E. 1994, Geophys. Res. Lett. 21, 2067, doi:10.1029/94GL01698

Hoyt, D.V., & Schatten, K.H. 1998, Solar Phys. 181, 491. doi:10.1023/A:1005056326158

IAU 2015, *"Corrected Sunspot History Suggests Climate Change since the Industrial Revolution not due to Natural Solar Trends"*, International Astronomical Union, Press Release IAU 108. http://www.iau.org/news/pressreleases/detail/iau1508/

IPCC 2013, *"Climate Change 2013: The Physical Science Basis"*. Contribution of Working Group I to the Fifth Assessment Report of the Intergovernmental Panel on Climate Change ed. Stocker, T.F. et al., Cambridge University Press, Cambridge, United Kingdom and New York, NY, USA, 1535 pp. http://www.ipcc.ch/report/ar5/wg1/

Ineson, S., Maycock, A.C., Gray, L.J., et al. 2015, Nature Communications, 6, Article # 7535, doi: 10.1038/ncomms8535

Johnson, R.W. 2011, Astrophys. Space Sci. 332, 73. doi: 10.1007/s10509-010-0500-1

Jones, G.S., Lockwood, M. and Stott, P.A. 2012, J. Geophys. Res. (Atmos.), 117, D05103, doi:10.1029/2011JD017013

Kopp, G., Krivova, N., Wu, C.J., & Lean, J. 2016, Solar Phys., in press, doi: 10.1007/s11207-016-0853

Leussu, R., Usoskin, I. G., Arlt, R., & Mursula, K. 2013, Astronomy and Astrophysics, 559, A28, doi: 10.1051/0004-6361/201322373

Lockwood, M. 2003, J. Geophys. Res., 108(A3), 1128, doi:10.1029/2002JA009431.

Lockwood, M. 2013, Living Rev. Solar Phys., 10, 4, doi: 10.12942/lrsp-2013-4. http://www.livingreviews.org/lrsp-2013-4.

Lockwood, M. & Owens, M.J. 2009, Astrophys J., 701(2), 964, doi:10.1088/0004-637X/701/2/964.

Lockwood, M., & Owens, M.J. 2014, J. Geophys. Res. Space Physics, 119, doi: 10.1002/2014JA019973.

Lockwood, M., Stamper, R. and Wild, M.N. 1999, Nature, 399, 437, doi:10.1038/20867.

Lockwood, M., Rouillard, A. P., Finch, I. D., & Stamper, R. 2006, J. Geophys. Res., 111, A09109. doi: 10.1029/2006JA011640

Lockwood, M., Barnard, L., Nevanlinna, H., Owens, M.J., Harrison, R.G., Rouillard, A.P., & Davis, C.J. 2013a, Annales Geophys. 31, 1957, doi:10.5194/angeo-31-1957-2013

Lockwood, M., Barnard, L., Nevanlinna, H., Owens, M.J., Harrison, R.G., Rouillard, A.P., & Davis, C.J. 2013b, Annales Geophys., 31, 1979, doi:10.5194/angeo-31-1979-2013

Lockwood, M., Owens, M.J. & Barnard, L.A. 2014a, J. Geophys. Res. Space Physics, 119 (7), 5172, doi: 10.1002/2014JA019970

Lockwood, M., Owens, M.J. & Barnard, L.A. 2014b, J. Geophys. Res. Space Physics, 119 (7), 5183, doi: 10.1002/2014JA019972

Lockwood, M., Nevanlinna, H., Vokhmyanin, M., Ponyavin, D., Sokolov, S. Barnard, L. Owens, M.J., Harrison, R.G., Rouillard, A.P., & Scott, C.J. 2014c, Annales. Geophys., 32, 367, doi:10.5194/angeo-32-367-2014

Lockwood, M,, Nevanlinna Barnard, L.,, H., Owens, M.J., Harrison, R.G., Rouillard, A.P., & Scott, C.J. 2014d, Annales. Geophys., 32, 383, doi:10.5194/angeo-32-383-2014

Lockwood, M., Scott, C.J., Owens, M.J., Barnard, L., Willis, D.M. 2016a,. Solar Phys., pre-published on-line, doi: 10.1007/s11207-016-0855-8

Lockwood, M., Scott, C.J., Owens, M.J., Barnard, L., Nevanlinna, H. 2016b, Tests of sunspot number sequences. 2. Using geomagnetic and auroral data. Solar Phys. (submitted).

Lockwood, M., Owens, M.J., Barnard, L., & Usoskin, I.G.: 2016c, Solar Phys. pre-published on-line, doi: 10.1007/s11207-015-0829-2

Lockwood, M., Owens, M.J., & Barnard, L.A. 2016d, Tests of sunspot number sequences: 4. Discontinuities around 1945 in various sunspot number and sunspot group number reconstructions, Solar Phys. (submitted).

Lockwood, M., Owens, M.J., & Barnard, L.A. 2016e, Comment on "Toward more reliable long-term indices of geomagnetic activity: Correcting a new inhomogeneity problem in early geomagnetic data" by L. Holappa and K. Mursula, J. Geophys. Res Space Phys., submitted.

Lynch, B.J., Gruesbeck, J.R., Zurbuchen, T.H. & Antiochos, S.K. 2005, J. Geophys. Res., 110, A08107, doi:10.1029/2005JA01113

Mackay, D.H., & Lockwood, M. 2002, Solar Phys., 209(2), 287-309, doi: 10.1023/A:1021230604497

Mackay, D.H., Priest, E.R., & Lockwood, M. 2002, Solar Phys., 207(2), 291-308, doi: 10.1023/A:1016249917230

Maycock, A.C., Ineson, S., Gray, L.J. et al. 2015, J. Geophys. Res. (Atmos.), 120, doi: 10.1002/2014JD022022

Nau, R. 2016, Statistical forecasting: notes on regression and time series analysis, http://people.duke.edu/~rnau/411home.htm

Owens, M.J. 2008, J. Geophys. Res., 113(A12), A12102, doi:10.1029/2008JA013589

Owens, M.J., & Crooker, N.U. 2006, J. Geophys. Res., 111, A10104, doi: 10.1029/2006JA011641

Owens, M.J., & Crooker, N.U. 2007, J. Geophys. Res., 112, A06106, doi: 10.1029/2006JA012159.

Owens, M.J., & Lockwood, M. 2012, J. Geophys. Res., 117, A04102, doi: 10.1029/2011JA017193.

Owens, M.J., Cliver, E., McCracken, K., Beer, J., Barnard, L.A., Lockwood, M., Rouillard, A.P., Passos, D., Riley, P., Usoskin, I.G. , & Wang, Y.-M. 2016, J. Geophys. Res., submitted

Satterthwaite, F.E. 1946, Biom. Bull. 2, 110. doi: 10.2307/3002019

Schwadron, N.A., Connick, D.E. & Smith, C.W. 2010, Astrophys. J., 722, L132–L136, doi: 10.1088/2041-8205/722/2/L132.

Sheeley Jr, N., Knudson, T. & Wang, Y.-M. 2001, Astrophys. J. Lett., 546(2), L131

Solanki, S., Schüssler, M., & Fligge, M. 2000, Nature, 480, 445, doi:10.1038/35044027.

Svalgaard, L. 2011, Proc. Int. Astron. Union 7, 27. Cambridge University Press, Cambridge, UK. doi:10.1017/S1743921312004590.

Svalgaard, L. 2014, Annales Geophys., 32 (6) 633, doi: 10.5194/angeo-32-633-2014

Svalgaard, L., & Cliver, E. W. 2007, *"Calibrating the sunspot number using "the magnetic needle",* CAWSES Newsletter, 4(1), 6-8.

Svalgaard, L., & Schatten, K.H. 2016, Solar Phys., pre-published on-line, doi: 10.1007/s11207-015-0815-8

Tlatov, A. & Ershov, V. 2014, Numerical processing of sunspot images using the digitized Royal Greenwich Observatory Archive, http://www.leif.org/research/SSN/Tlatov2.pdf

Usoskin, I.G. 2013, Living Rev. Solar Phys. 10, 1. doi:10.12942/lrsp-2013-1, http://www.livingreviews.org/lrsp-2013-1

Usoskin, I.G., Kovaltsov, G.A., Lockwood, M., Mursula, K., Owens, M.J., & Solanki, S.K.: 2016, Solar Phys., pre-published on-line, doi: 10.1007/s11207-015-0838-1

Vaquero, J.M., Gallego, M.C., Usoskin, I.G., &. Kovaltsov G.A. 2011, Astrophys. J., 731(2), L24, doi:10.1088/2041-8205/731/2/L24.

Vieira, L.E.A., & Solanki, S.K. 2010, Astrophys. J., 509(1), A100, doi: 10.1051/0004-6361/200913276.

Welch, B.L. 1947, Biometrika 34, 28, doi: 10.1093/biomet/34.1-2.28

Willis, D.M., Wild, M.N., & Warburton, J.S.: 2016, Solar Phys., pre-published on-line, doi: 10.1007/s11207-016-0857-7

Willis, D.M., Coffey, H.E., Henwood, R., Erwin, E.H., Hoyt, D.V., Wild, M.N., & Denig, W.F.: 2013a, Solar Phys. 288, 117. doi: 10.1007/s11207-013-0311-y

Willis, D.M., Henwood, R., Wild, M.N., Coffey, H.E., Denig, W.F., Erwin, E.H., & Hoyt D.V.: 2013b, Solar Phys. 288, 141. doi: 10.1007/s11207-013-0312-x

Wolf, R. 1861, Mitth. über die Sonnenflecken, 12, doi: 10.3931/e-rara-3058

Wolf, R. 1873, Astronomische Mittheilungen, Nr. 30. Vierteljahrsschrift der Naturforschenden Gesellschaft in Zürich, 1.

Yashiro, S., Gopalswamy, N., Michalek, G., St. Cyr, O.C., Plunkett, S.P., Rich, N.B., & Howard, R.A. 2004, J. Geophys. Res., 109, A07105, doi:10.1029/2003JA010282.

**Table 1.** Comparison of metrics for the fits shown in figure 3.

|  | $R_C$ | $R_{BB}$ | $R_{UEA}$ | $R_G$ | $R_{ISNv2}$ | $R_{ISNv1}$ |
|---|---|---|---|---|---|---|
| $r$ | 0.9091 | 0.9086 | 0.8959 | 0.8785 | 0.9116 | 0.9034 |
| $S_r$ (%) | 99.9986 | 99.9995 | 99.9991 | 99.9933 | 99.9995 | 99.9979 |
| $\Delta$ ($10^{15}$ Wb) | 0.0350 | 0.0310 | 0.0315 | 0.0399 | 0.0308 | 0.0364 |
| $\Delta_P$ ($10^{15}$ Wb) | 0.0483 | 0.0351 | 0.0547 | 0.0541 | 0.0356 | 0.0472 |
| $\Delta_T$ ($10^{15}$ Wb) | 0.0547 | 0.0508 | 0.0569 | 0.0559 | 0.0556 | 0.0610 |

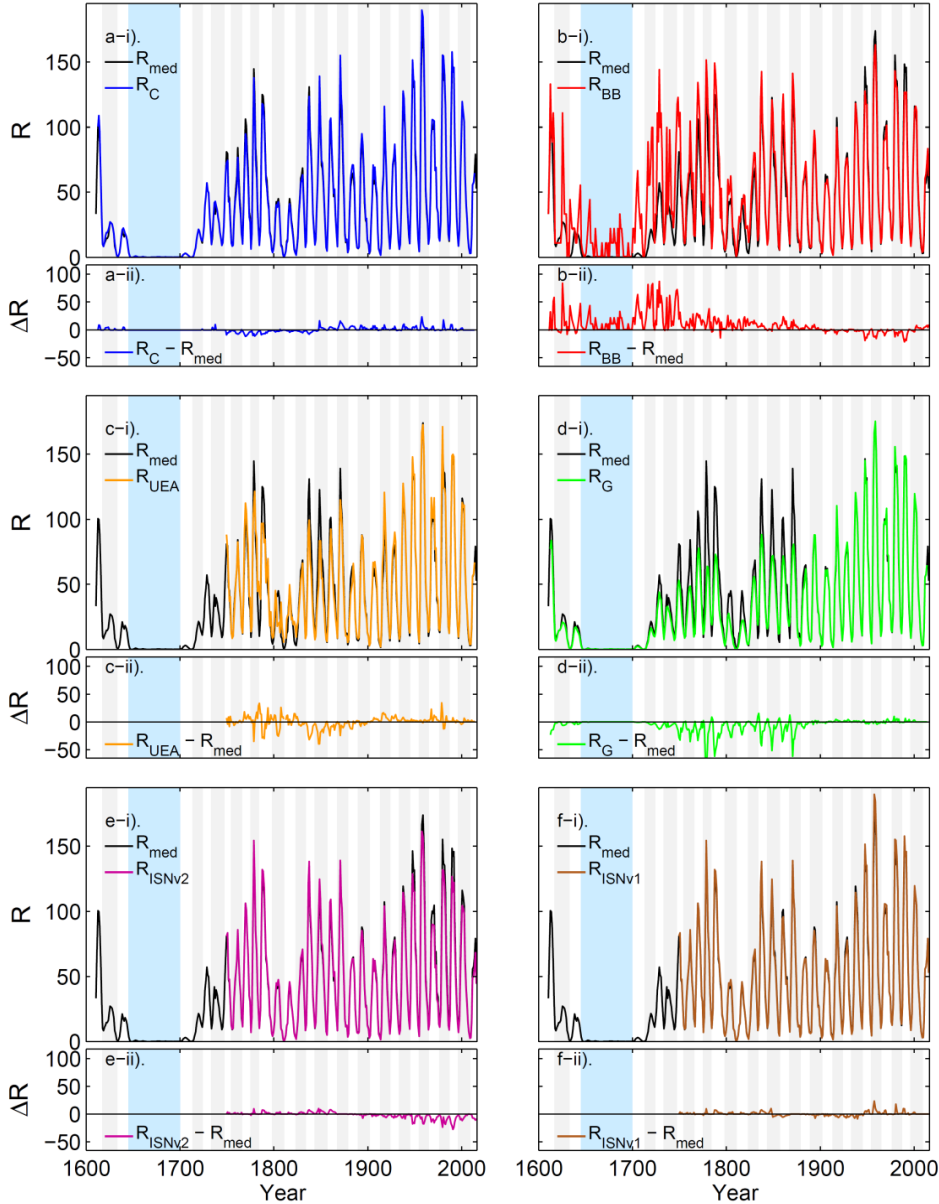

**Figure 1**. The various sunspot number sequences studied in this paper. Each is here compared to the median of all available sequences in that year (which vary in number from 3 in 1650 to 6 in 2015), $R_{med}$, shown in black in each panel. Grey and white vertical bands define, respectively, odd- and even-numbered sunspot cycles (from minimum to minimum) and the cyan band is the Maunder minimum. (a-i) The corrected sunspot number, $R_C$ (in blue), proposed by *Lockwood et al.* (2014). (b-i) The "backbone" group number reconstruction, $R_{BB}$ (in red), of *Svalgaard and Schatten* (2016). (c-i) The group number derived by *Usoskin et al.* (2016), $R_{UEA}$ (in orange). (d-i) The *Hoyt and Schatten* (1998) group number, $R_G$ which has been extended to 2015 using the SOON dataset, as calibrated against $R_G$ by *Lockwood et al.* (2015). (e-i) Version 2 of the international sunspot number, $R_{ISNv2}$, introduced by SIDC (see text) in July 2015 (in purple) (*Clette et al.*, 2014). (f-i) Version 1 of the international sunspot number, $R_{ISNv1}$ that was issued by SIDC until July 2015 (in brown). To help identify the differences, the lower panels in each pair show the difference between each and $R_{med}$ (so a-ii shows $R_C - R_{med}$, etc.).

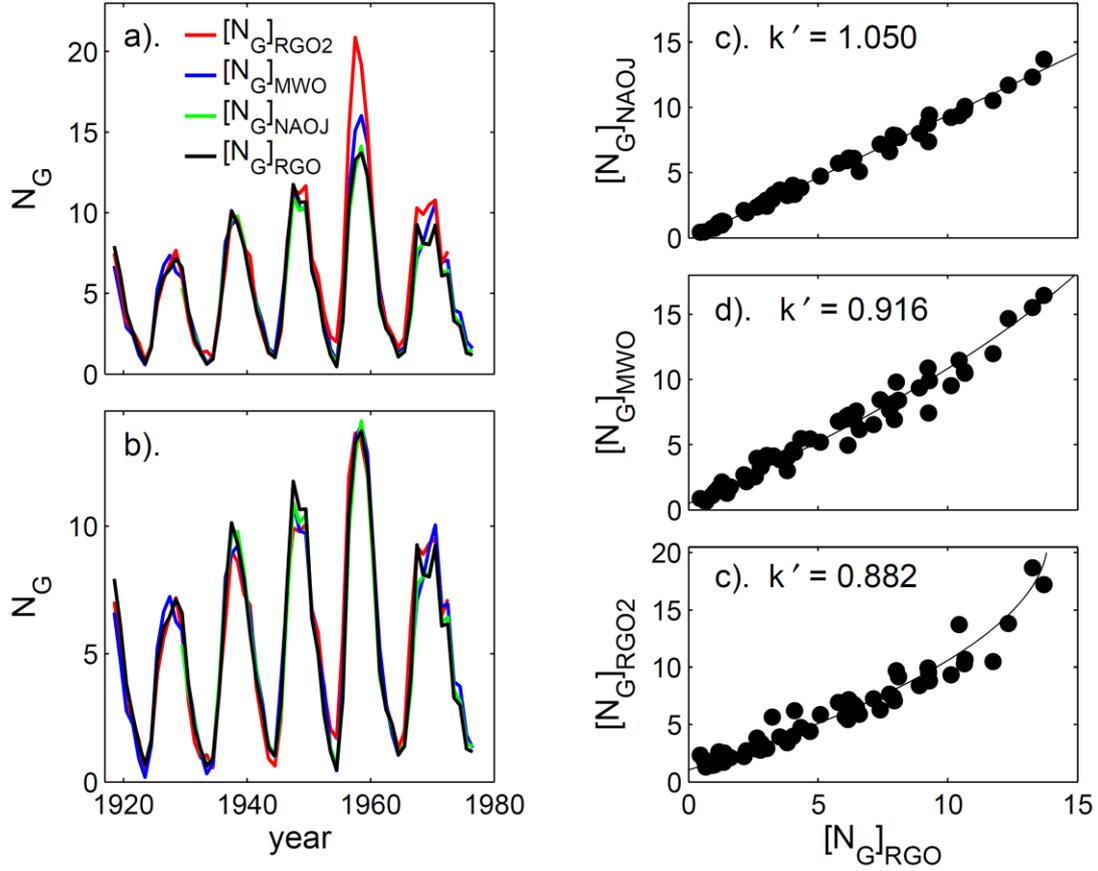

**Figure 2**. Comparison of sunspot group number data from various observers. The time series in (a) have been scaled to the standard RGO dataset ($[N_G]_{RGO}$, in black) over 1920-1945 using linear regression: from Mount Wilson Observatory ($[N_G]_{MWO}$, in blue), from the Solar Observatory of the National Astronomical Observatory of Japan ($[N_G]_{NAOJ}$, in green), and from the auto-scaled RGO photographic plates ($[N_G]_{RGO2}$, in red). (b) The same data series as in (a), scaled using a $2^{nd}$-order polynomial fit to $[N_G]_{RGO}$ over 1920–1976. (c)–(e) scatter plots and $2^{nd}$-order polynomial fits for the interval 1920–1976 as a function of $[N_G]_{RGO}$ for: (c) $[N_G]_{NAOJ}$; (d) $[N_G]_{MWO}$; and (e) $[N_G]_{RGO2}$.

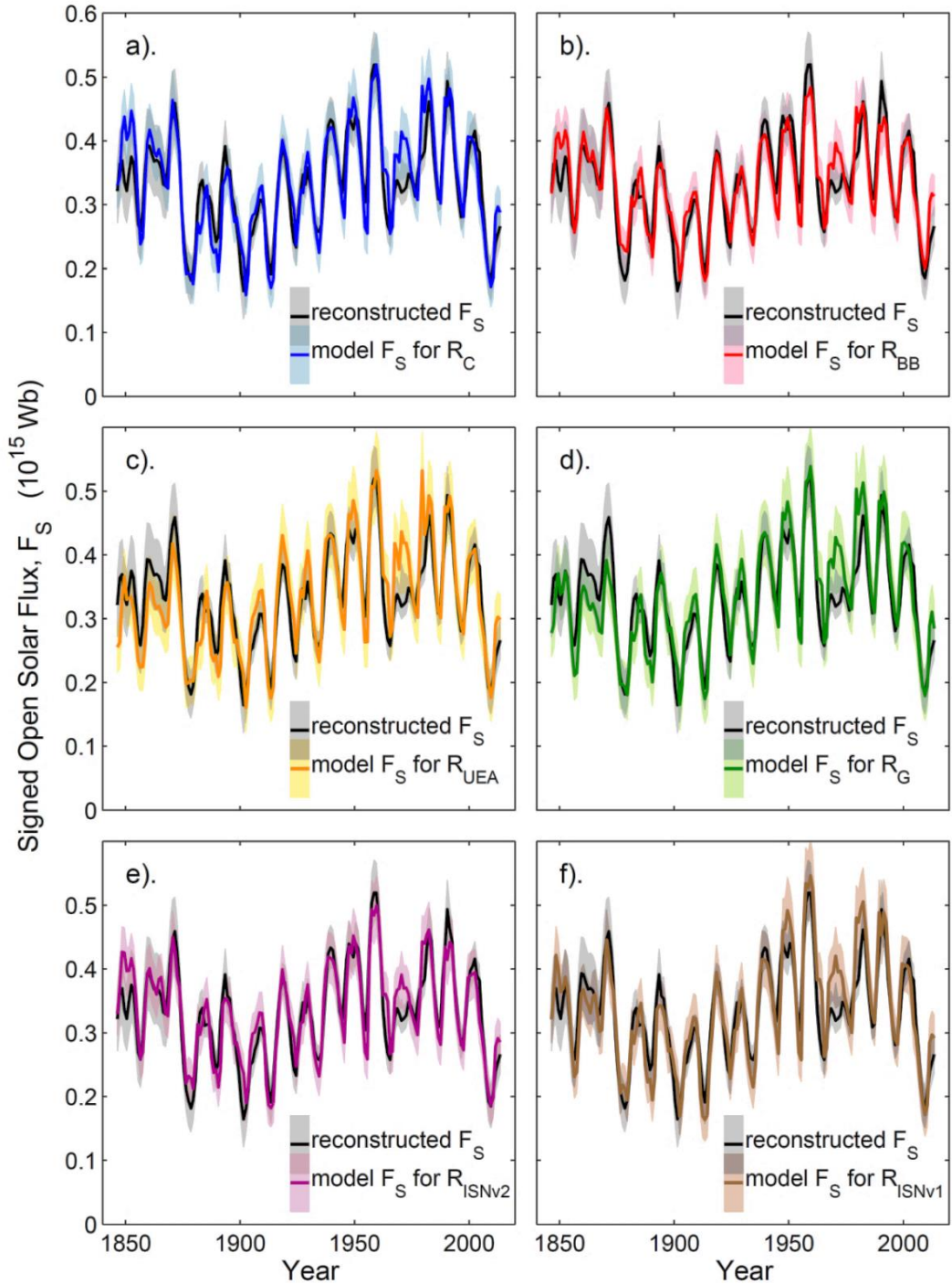

**Figure 3.** Comparisons of the reconstruction by *Lockwood et al.* (2014b) of the signed open solar flux, $F_S$, from 4 different pairings of geomagnetic activity indices (in black with its $\pm 1\sigma$ uncertainty band shown in grey) and the modelled open solar flux using the model of *Owens and Lockwood* (2012) using the sunspot number sequences shown in figure 1 to quantify the emergence of open solar flux:- (a) for $R_C$ (in blue); (b) $R_{BB}$ (in red); (c) $R_{UEA}$ (in orange); (d) $R_G$ (in green); (e) $R_{ISNv2}$ (in purple) and (f) $R_{ISNv1}$ (in brown). The $\pm 1\sigma$ uncertainty band in each modelled $F_S$ variation is shown in a lighter shade of the line colour in each case and the darker shade shows the overlap of the uncertainty bands of the modelled and reconstructed $F_S$.

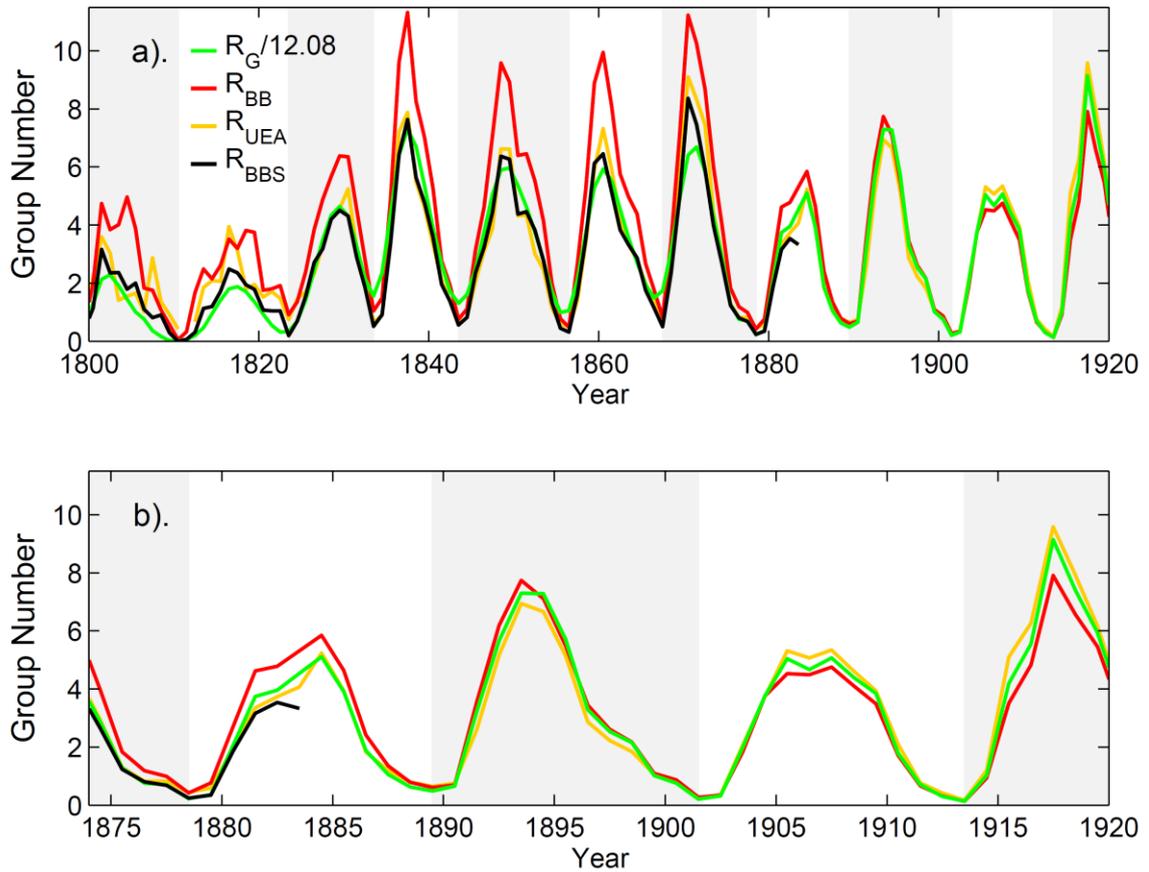

**Figure 4**. Variations in annual means in and between the intervals covered by the Schwabe and Wolfer data. The green lines show the *Hoyt and Schatten* (1998) group number, $R_G$; the red line is the "backbone" reconstruction of *Svalgaard and Schatten* (2016), $R_{BB}$; the orange line is the group number reconstruction of *Usoskin et al.* (2016), $R_{UEA}$; the black line is the "Schwabe backbone" generated by *Svalgaard and Schatten* (2016), $R_{BBS}$, which they multiply by 1.48 to obtain $R_{BB}$, that being the factor that they derived from linear regression (assuming proportionality) of the Schwabe and Wolfer backbones over 1861−1883. Grey and white vertical bands define, respectively, odd- and even-numbered sunspot cycles. (a) covers the interval 1800−1920 and (b) shows 1874−1920 in greater detail.

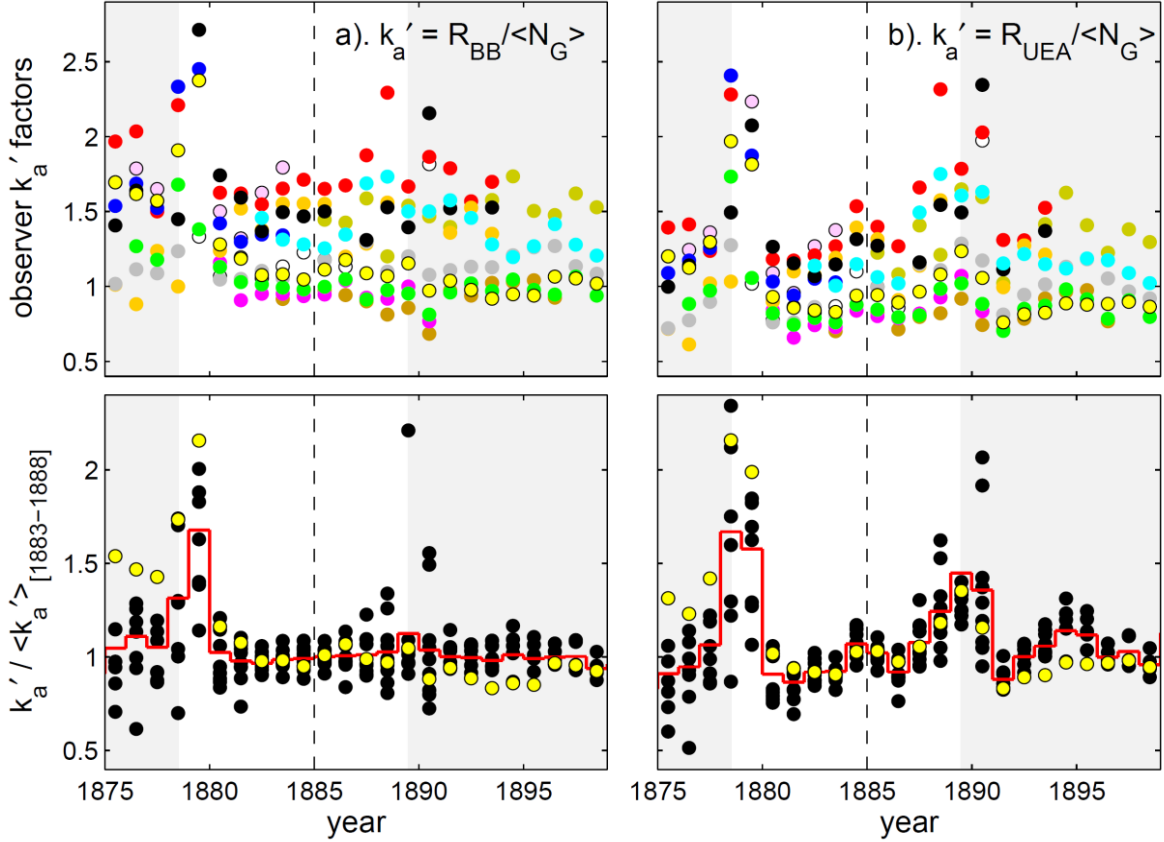

**Figure 5**. Analysis of the variations of annual group number observer factors, $k_a'$, for various observers making observations in the interval covered by the 20[th] century RGO data: (a) for $R_{BB}$ (i.e., $k_a' = R_{BB}/\langle N_G \rangle$, where $\langle N_G \rangle$ is the annual mean of the sunspot group counts recorded by each observer); (b) for $R_{UEA}$ (i.e., $k_a' = R_{UEA}/\langle N_G \rangle$). Observers are: (orange) Spörer; (red) Wolf (using the small telescope); (blue) Schmidt; (grey) Tacchini; (pink) Weber; (green) Wolfer; (mauve) Rico; (black) Moncalieri ; (brown) Merino; (olive) Konkoly; (white) Dawson; (yellow) RGO; and (cyan) Winkler . The lower panels show the $k_a'$ values normalised by dividing by their average values over a reference period of 1883−1888: the yellow dots are for the RGO data and the red histogram shows the mean of all normalised values, excluding the RGO data. The vertical dashed line is 1885 when Cliver and Ling (2016) infer a discontinuity in the RGO data.

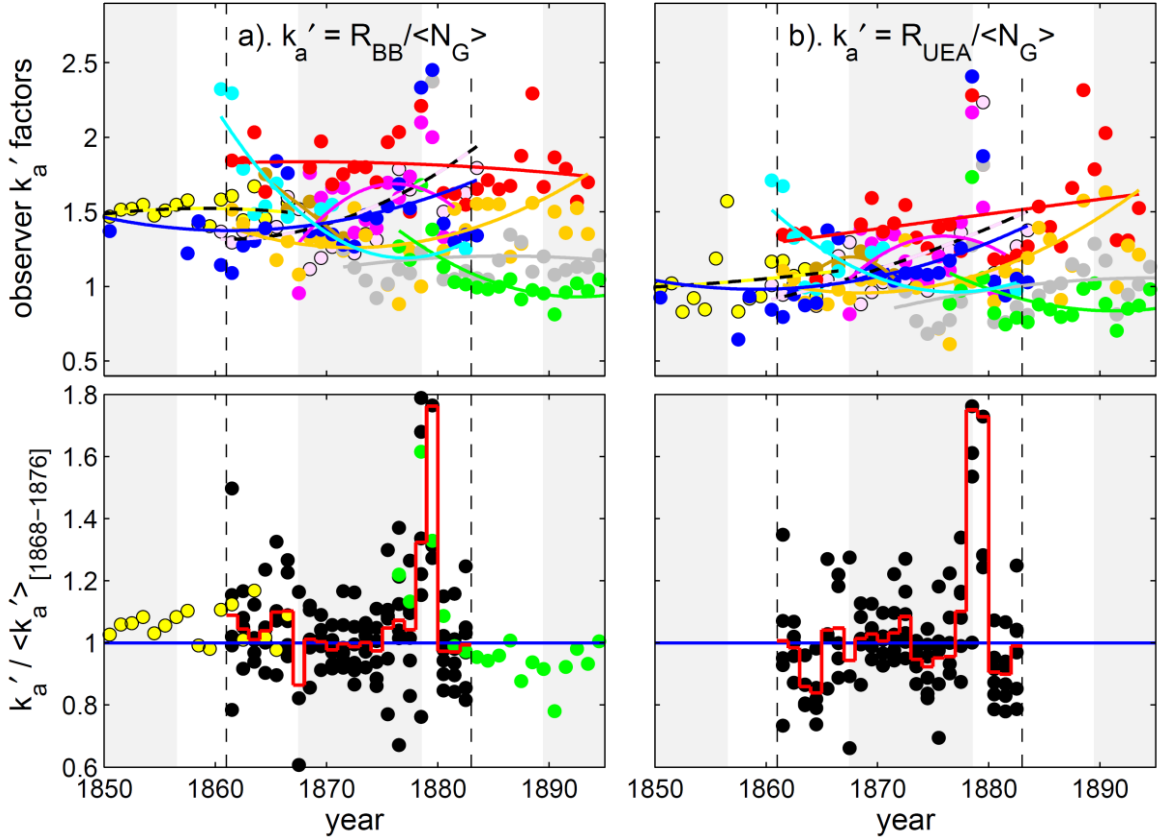

**Figure 6**. Same as figure 5 for all observations used to join the Schwabe and Wolfer backbones: (a) for $R_{BB}$ (i.e., $k_a' = R_{BB}/\langle N_G \rangle$, where $\langle N_G \rangle$ is the annual mean of the sunspot group counts recorded by each observer); (b) for $R_{UEA}$ (i.e., $k_a' = R_{UEA}/\langle N_G \rangle$). Observers are: (orange) Spörer; (red) Wolf (using the small telescope); (blue) Schmidt; (grey) Tacchini; (mauve) Leppig; (pink) Weber; (cyan) Howlet; (brown) Meyer; (yellow) Schwabe; and (green) Wolfer. In addition to the annual $k_a'$ values, the upper panels here show second-order polynomial fits to the points for each observer to demonstrate the variations. The vertical dashed lines delineate the interval over which the Schwabe and Wolfer backbone were correlated in the daisy-chaining used to generate $R_{BB}$. The lower panels show the $k_a'$ values normalised by dividing by their average values over a reference period of 1868–1876. The red histogram shows the mean of all normalised values. In the lower panel of (a), the yellow and green dots are the data of Schwabe and Wolfer, intercalibrated using the red histogram. Note that the data shown here were used to intercalibrate the data of Schwabe and Wolfer in the construction of $R_{BB}$ but were not used for that intercalibration in the generation of $R_{UEA}$.